\documentclass[USenglish,10pt,authoryear,twocolumn]{article}
\usepackage[top=.75in,left=2.25cm,bottom=1in,right=2.25cm]{geometry}
\usepackage{geometry}
\usepackage{graphicx}
\usepackage
{natbib}

\usepackage{epstopdf}
\usepackage{amsmath}
\usepackage{amssymb}
\usepackage{multirow}
\usepackage[utf8]{inputenc}
\usepackage{textcomp}
\usepackage[usenames,dvipsnames]{color}
\usepackage[bookmarks, colorlinks, breaklinks, pdfauthor={Taramasco, C., Cointet, J.-Ph., Roth, C.}]{hyperref}
\hypersetup{linkcolor=MidnightBlue,citecolor=MidnightBlue}
\usepackage{enumerate}

\newcommand{\cameraready}[1]{#1}
\newcommand{\pics}[1]{}

\newcommand{\Aset}{\mathcal{A}}
\newcommand{\Cset}{\mathcal{C}}
\newcommand{\hyperA}[1]{\mathfrak{A}_{#1}}
\newcommand{\hyperC}[1]{\mathfrak{C}_{#1}}
\newcommand{\hyperE}[1]{\mathfrak{E}_{#1}}
\newcommand{\hypere}{\mathfrak{e}}
\newcommand{\Aproj}[1]{{#1}^{\Aset{}}}
\newcommand{\Cproj}[1]{{#1}^{\Cset{}}}

\newcommand{\papertitle}{Academic team formation as evolving hypergraphs}
\title{\papertitle{}}

\newcommand{\iscpif}{ISC-PIF (Institut des Systèmes Complexes -- Paris-Île-de-France). 56, rue Lhomond, 75005 Paris, France.}
\newcommand{\crea}{CREA (CNRS/Ecole Polytechnique, France). CREA/ENSTA, 45 Bd Victor, 75015 Paris, France}
\newcommand{\decom}{DECOM (Universidad de Valparaiso, Chile). Avenida Gran Bretaña, 1091 Playa Ancha. Valparaiso, Chile}
\newcommand{\ifris}{INRA SenS (Sciences en Société) - IFRIS. 5, Bd Descartes. 77420 Champs-sur-Marne, France}
\newcommand{\cams}{CAMS (CNRS/EHESS, France). EHESS/CNRS, 54 Bd Raspail, 75006 Paris, France}
\newcommand{\cress}{CRESS (U. Surrey, GB). University of Surrey, Guildford GU2 7XH, United Kingdom
\smallskip\newline{\em Emails:}\newline
roth@ehess.fr (\emph{corresponding author})\newline Carla.Taramasco@polytechnique.edu\newline Jean-Philippe.Cointet@polytechnique.edu}

\author{Carla Taramasco\footnote{\iscpif} $\,^,$\footnote{\crea} $\,^,$\footnote{\decom} \and Jean-Philippe Cointet\footnotemark[1] $\,^,$\footnotemark[2] $\,^,$\footnote{\ifris} \and Camille Roth\footnotemark[1] $\,^,$\footnotemark[2] $\,^,$\footnote{\cams} $\,^,$\footnote{\cress}}
\date{\small \sc {[}preprint --- paper to appear in \href{http://www.kluweronline.com/issn/0138-9130/contents}{scientometrics}{]}}
\newcommand{\paperabstract}{This paper quantitatively explores the social and socio-semantic patterns of constitution of academic collaboration teams. To this end, we broadly underline two critical features of social networks of knowledge-based collaboration: first, they essentially consist of group-level interactions which call for team-centered approaches. Formally, this induces the use of {\em hypergraphs} and $n$-adic interactions, rather than traditional dyadic frameworks of interaction such as \emph{graphs}, binding only pairs of agents. Second, we advocate the joint consideration of structural and semantic features, as collaborations are allegedly constrained by both of them.
Considering these provisions, we propose a framework which principally enables us to empirically test a series of hypotheses related to academic team formation patterns. In particular, we exhibit and characterize the influence of an implicit group structure driving recurrent team formation processes. On the whole, innovative production does not appear to be correlated with more original teams, while a polarization appears between groups composed of experts only or non-experts only, altogether corresponding to collectives with a high rate of repeated interactions.}

\begin{document}
\maketitle
\begin{abstract}
\paperabstract{}
\end{abstract}


\section{Introduction}

The mechanisms of academic collaboration are the focus of a long and established tradition of research \citep{katz:what}, from qualitative studies on cooperation and co-optation behaviors \citep{cran:soci,chub:conc,lato:labo} to {more} quantitative approaches \citep{beav:stud,beav:coll,meli:stud}. The latter includes network-based studies, which are generally aiming at understanding the structural determinants and patterns of collaboration  \citep{mull:deve,newm:str12,bara:evol,mood:strusoci,wagn:netw,leah:rese}.
In this case, the quantitative formal framework of choice is the social network of dyadic interactions, addressing questions related to how ego-centered characteristics, \emph{in the broad sense}, influence the likelihood of being involved in a collaboration.
 
\subsubsection*{The Team Level and Networks}
Network studies, specifically in the context of scientific collaboration, indeed often focus on the level of the individual 
in spite of a large amount of work on the question of group cohesiveness \citep{lott:grou,boll:perc,frie:soci}.
There are wider implications of this focus on the ego-centered level:
	\begin{itemize}
	\item By aiming at describing individual behavioral patterns, this perspective may overlook the influence of characteristics expressable at the meso-level of the team itself. In particular, by focusing on dyadic interactions and  relational patterns between ego and alter(s), the presence of ego in a given collaboration is interpreted as a function of the characteristics of ego and those of alter(s), and of the characteristics of the various dyads between ego and alter(s).

	\item Further, the creation of a group results from a complex agreement and arrangement between all its members, who jointly decide to collaborate. As such, even when assuming that the behavior of ego may depend on non-dyadic, team-level characteristics, interpreting team formation processes as a sum of individual rationalities may oftentimes seem difficult, or irrelevant. Put differently, there are regularities in team formation processes which are difficult to ascribe specifically back to individuals; it may appear more natural and consistant to appraise the underpinnings of group formation at the group level.\footnote{Note that what we call a ``team'' here actually relates to a group that is involved in the production of an academic paper, i.e. the team of coauthors that produces it; it does not correspond to the more or less explicit notion of team that may exist in some research labs.}
	
\end{itemize}
	
To sum up, when dyadic frameworks are involved, collaboration teams are appraised under the lens of multiple one-to-one interactions. It should be no surprise: social network literature is itself overwhelmingly concerned with dyadic links.
However, a sizeable portion of sociology, starting with \cite{simm:pers}, has long been concerned by wider frameworks of interactions, or so-called ``social circles'', which some authors have formalized to take directly into account non-dyadic relationships: \cite{brei:dual,brei:soci}, for instance, proposed to use bipartite graphs to represent and analyze ties between actors and social groups. Focusing on the group-level, \cite{ruef:stru} quantitatively examined the contribution of several factors including gender, status, or ethnicity, in the preferential constitution of business founding teams. In a review study, \cite{free:find} explored various approaches previously adopted in mathematical sociology to model two-mode data in order to account for the presence of subsets of people participating altogether in (subsets of) identical events.

In this respect, it therefore first appears that academic collaboration choices and dynamics should be characterized by investigating the meso-level of team formation. More precisely, it should be fruitful to focus on \emph{teams} rather than pairs of agents interacting together, thus advocating the use of \emph{hypergraphs} or bipartite graphs rather than traditional frameworks based on graphs. 
Hypergraphs indeed feature \emph{hyperlinks} which connect arbitrary numbers of agents, while graphs feature \emph{links} which connect only \emph{pairs} of agents. 
In other words, considering hypergraphs prevents making the superfluous and plausibly debatable assumption that teams are equivalent to complete subgraphs featuring one-to-one interactions between all its members (i.e. assuming for instance that a triad is equivalent to three dyads).

\subsubsection*{Hybrid Networks of Actors and Concepts} 

Secondly, collaboration massively depends on cognitive properties, in particular some cognitive fit between team members, as agents plausibly compose teams in order to gather complementary competences. For instance, some economic models of knowledge creation consider matching rules based on the similarity of agent profiles, as elements of a vector space, to explain economic network structure \citep{cowajonazimm:join}.
In other words, equal attention should be given to social and semantic features, which are traditionally left apart in the literature, although the existence of homophily-driven interactions has been underlined in numerous works \citep{mcph:bird}.

Our main hypothesis is that one cannot correctly understand the underlying social processes if both social and semantic dimensions of, e.g., scientific activity, are not considered as two interdependent dynamics\cameraready{ \citep{roth:phd,roth:soci}}. Going further, we construe scientific dynamics as made of groupings of both agents and concepts: the \emph{epistemic} dynamics, \hbox{i.e.} the collective scientific knowledge construction, is made of events which simultaneously involve compounds of actors and concepts. In line with the program introduced by \cite{call:some}, we will appraise scientific dynamics as  made of constant reconfiguration and re-negotiation of collectives of both humans and non-humans. 


In this respect and more broadly, in addition to focusing on teams, we thus  advocate the enrichment of the notion of team by \emph{considering teams as joint groupings of both agents and semantic items}.

\subsubsection*{Knowledge-based teamwork}

The interest in the social epistemology of academic communities also has a broader reach. As a knowledge production arena, science is indeed likely to share features found in other collaborative knowledge creation contexts. 
\begin{enumerate}[(i)]
\item {\em Collaboration in knowledge production systems.}

This issue may shed light, to some extent,  on the interaction processes underlying, broadly, collaborative knowledge production. 
These contexts indeed define a particularly common class of social networks of collaboration, where  agents  jointly and collectively interact for purposes of knowledge production, in the broad sense. This encompasses activist groups and political epistemic communities \citep{rugg:inte,haas:intr}, scientific communities \citep{beav:stud,laba:inte,jone:mult,stok:scie,leah:rese} and more specifically research projects \citep{lare:stru,lare:netw}, open-source development communities \citep{kogu:open} and discussion lists and forums \citep{cons:kind,wels:visu}, wiki platform-mediated communities \citep{brya:beco,levr:wiki}, 
artists gathering for a theater performance \citep{uzzi:coll} or making a movie \citep{faul:shor,rama:self}, board members making collective decisions \citep{davi:corp}.

\item {\em Collaboration in teams.}

This kind of relatively autonomous collaboration mode has to be understood in a context where traditionally vertical and hierarchical organizations have recently been functioning in increasingly horizontal and networked ways \citep{powe:neit,mile:orga,sman:natu}. This contemporary so-called ``network governance'' involves dynamic coalitions of actors both at organizational and individual levels, increase of teamwork and frequent group reconfigurations \citep{jone:gene}.
This shift is particularly sensible in contexts where agents are relatively free to group to form casual alliances and where collaboration sometimes appears to be self-organized. 
\end{enumerate}

In this respect, science appears to be a  prototypical case of such teamwork-based systems \citep{beav:coll,adam:scie,wuch:incr} --- scientific knowledge production essentially involves events where researchers jointly work to manipulate and introduce concepts. It is additionally one of the most accomplished context of \emph{knowledge-based} collaboration, as well as one of the most explicit, by its very stigmergic\footnote{{``Stigmergic'': that is, leaving \emph{traces} susceptible to guide the work of others.  For an extensive discussion of this notion, see \cite{kars:comb}.}} nature: papers indeed constitute a concrete, often public instance of these gatherings and therefore provide an opportunity to understand the impact of these collaborations on the dynamics of science. 
On the empirical side, we thus rely on large bibliographic databases. 

{As such, our approach does not pretend to embrace the whole complexity of knowledge-intensive organizations, in particular the intricate co-evolutionary processes existing between formal organizations and more local team-based and individual-based decisions \citep{laze:catc}. However, the metholodogy we propose is able to shed some original light on portions of the dynamics of these knowledge production systems. }

\bigskip\bigskip
The paper is organized as follows: in Sec.~\ref{sec:framework}, we present the framework and   support several hypotheses on socio-semantic team-based collaboration, Sec.~\ref{sec:protocol} introduces the protocol and methods, while Sec.~\ref{sec:results} presents the results, which we then discuss in light of the initially proposed hypotheses.

\section{Framework}\label{sec:framework}

As follows from the introduction, we hence argue that two features are key in extending the understanding of, one hand, collaboration networks, and on the other hand and additionally, knowledge production networks: 
\begin{enumerate}
\item Group effects underlie and partially determine dyadic interactions: affiliation to teams of collaboration, membership in identical epistemic communities, for instance, structure and influence the very formation of these interactions.


\item In the case of social networks of knowledge, these underlying groups are both social (work communities) and semantic (epistemic communities). 
In particular, the choice of collaboration partners is likely to highly depend on cognitive similarity.
\end{enumerate}


\bigskip
More to the point, in terms of strictly social and strictly semantic associations, we first aim at checking the following simple hypotheses, by comparing what happens empirically with what would have happened if teams had been formed strictly by chance (\hbox{i.e.} by comparing empirical teams with a null-model featuring random compositions of teams).
\begin{enumerate}[\quad\bf (H1).]
\item Teams with a high rate of interaction repetition should be more likely, as could be expected because of social cohesion \citep{boll:perc,mcph:cohe,frie:soci} or organizational constraints \citep{rodr:rela}.

\item Teams where a high proportion of concepts are repeatedly associated should be more likely --- as assumed by co-word analysis \citep{call:mapp,noyo:moni}, where frequent associations of terms are supposed to define conceptual cores and field boundaries.

\item Papers with a higher semantic originality (\hbox{i.e.} new association of concepts) should be those where there is a higher number of new interactions.\footnote{As \citet[p.414]{call:scie} sums up from the existing literature,
\begin{quote}
``The more numerous and different these heterogeneous collectives are, the more the reconfigurations produced are themselves varied''
\end{quote}}
Put differently, as suggested by social and semantic repetitions assumed by H1 and H2, teams with a high number of repeated interactions should tend to produce papers that have smaller semantic/topical originality; which in some sense belong to a narrower subfield of research \citep{leah:rese}.
\end{enumerate}

Then, we appraise the socio-semantic composition of teams. We more precisely focus on the distinction between agents who are already familiar with some concepts involved in the interaction, and those who are not. This approach will more broadly inform us about the cognitive specialization of teams. 
\begin{enumerate}[\bf (HI).]
\item Because of both scientific specialization \citep{chub:conc} and homophily \citep{mcph:bird,stok:scie}, teams gathering around a given topic should generally involve more individuals knowledgeable about this given topic.

\item Teams with a balanced composition of experts in a given field should produce more innovation \citep{anco:demo}, which in terms of networks could be translated into:
\begin{itemize}
	\item more semantic originality, i.e. novel associations of concepts,
	\item more social originality, i.e. novel interactions between agents.
\end{itemize}
\end{enumerate}

\section{Protocol and methods}\label{sec:protocol}

In line with this focus on socio-semantic aspects, we will thus endeavor at exhibiting how new teams are formed by considering both social and conceptual past acquaintances of scientists involved in new collaborations. 
We will concretely describe the semantic dimension in terms of attributes qualifying topics of interest of authors and the social dimension as structural and relational properties in the dynamic collaboration network --- which altogether will enable us to confirm or refute the previous set of hypotheses. 

\subsection{Datasets}

Our empirical analysis focuses on collaboration databases, which reveal a large part of the underlying collaboration activity, including social links between individuals or conceptual acquaintances of each individual (\hbox{i.e.} details regarding which topics which agents are familiar with). 
These datasets provide temporal information on teams, gathering agents and the topics they work on, assuming that topics are described by the very terms used in paper abstract. For each dataset, we focus on a set of no more than a hundred of relevant terms. These terms are selected with the help of an expert of the corresponding field and are such that they appropriately cover the most significant topics of each field. \medskip

We use the following datasets, defined either from a semantic perspective (using \hbox{e.g.} field names) or from a social perspective (using \hbox{e.g.} scientific assemblies), and involving both large and small communities:
\begin{enumerate}
\item Embryologists working within a given and well-determined subfield --- the zebrafish, on a period of 20 years (1985--2004).
Data was extracted from the publicly available database \emph{Medline}, which eventually yields a dataset of $6,145$ articles ($13\,084$ authors, $71$ word classes).
\item Scientists working on rabies from the same kind of \emph{MedLine} extraction as for zebrafish embryologists --- the observed period spans from 1985 to 2007.
This ends up with $4\,648$ events ($9\,684$ authors, $70$ word classes).
\item Scientific committee members for JEMRA meetings\footnote{Joint FAO/WHO Expert Meetings on Microbiological Risk Assessment, http://www.fao.org/ag/agn/agns/jemra\_index\_en.asp}: this dataset includes the publications of an initial set of 168 scientists involved in these meetings, gathered from 1985 to 2007.
This ends up with $5\,893$ papers ($15\,375$ authors, $69$ word classes).
\item Scientific committees members for JECFA meetings\footnote{Joint FAO/WHO Expert Committee on Food Additives, http://www.who.int/ipcs/food/jecfa}: similarly, publications of an initial set of 178  scientists are gathered from 1985 to 2007.
This ends up with $8\,685$ papers ($21\,195$ authors, $85$ word classes).
\end{enumerate}

\subsection{Hypergraph-based definitions}

Now, {these agents and concepts \emph{formally} define an evolving hypergraph where each article is a hybrid hyperlink gathering both authors and the topics involved in the collaboration}, as partly exemplified by Fig.~\ref{fig:hypergraph}. 

In what follows, we describe comprehensively our formal framework (Sec.~\ref{sec:objects}), which, basically, allows us to gather both agents and concepts in a dynamic setting and to define which agents are new, or not (\emph{newcomers} vs. \emph{veterans}), which concepts are new, or not (\emph{novelties} vs. \emph{standards}), and which agents have used which concepts in the past, or not (\emph{neophyte} or \emph{experts}).

Building upon these definitions, we will then propose a series of \emph{hypergraphic} measures (Sec.~\ref{sec:measures}) --- that is, measures at the level of teams, or non-dyadic measures --- which cover the proportion of experts in a given collaboration (\emph{expertise ratio}) and the originality of participants in a team (\emph{hypergraphic repetition}, \hbox{i.e.} describing to what extent a team does gather agents, or concepts, which were jointly associated, at the team-level, in previous periods). 
For instance, a team with an expertise ratio of one will be such that all agents are experts; a team with a hypergraphic repetition of one, in terms of agents, will be such that all its agents will have \emph{altogether} previously collaborated (it is zero in case none of the agents have previously been associated).

Then, we present a methodology (Sec.~\ref{sec:model}) for computing how much the empirical data diverges from a random setting with a comparison between the actual observed data and a uniform \emph{null-model of hypergraph evolution}. Put simply, we will appraise how much teams with, \emph{e.g.}, a given hypergraphic repetition ratio, are  forming significantly more often than could be expected by chance. This latter tool will be the cornerstone of the empirical testing of hypotheses 1-2-3 \& I-II.

\subsubsection{Objects}\label{sec:objects}

\subsubsection*{Hypergraphs.}

Formally, a \emph{hypergraph} features \emph{nodes} and \emph{hyperlinks}, which describe $n$-adic interactions among any subset of nodes. It is therefore a generalization of the notion of graph whose links only describe dyadic interactions, i.e. between pairs of nodes. As such, any hyperlink corresponds to any grouping of agents and any kind of social circle: it may describe social events, organizations, families, teams, etc. A hypergraph is also isomorphic to a bipartite graph, where agents on one side are connected to various affiliations, groups or events on the other side; as such a structure which reifies the duality of social groups \citep{brei:dual,free:find}. See Fig.~\ref{fig:hypergraph}.

Beyond the simple observation of the structure of such networks, several studies have endeavored at reconstructing structural properties typically induced by the hypergraphic setting --- namely, that agents interact within groups of some sort  --- rather than using dyadic interactions only: 
in this direction \cite{newm:rand,rama:self,guim:team}, \emph{inter alia}, examine the structure of a social network whose dyadic links stem from teams --- team composition is first empirically appraised then stylized and used as a basis for what essentially is a clique addition process. In these models however the focus remains on dyadic relationships or dyadic interaction behaviors, rather than truly hypergraphic measures.

In contrast, the focal level of analysis of the present study is the hypergraph and its hyperlinks.


\subsubsection*{Epistemic hypergraphs.}

To bind the social and semantic aspects, we introduce the notion of \emph{epistemic hypergraph} $\mathfrak{E}_t$ using:
\begin{enumerate}[(i)]
\item a set of agents $\Aset{}$,
\item a set of concepts $\Cset{}$,
and 
\item the epistemic hypergraph itself \hbox{$\hyperE{}\subseteq\mathcal{P}(\Aset{}\cup\Cset{})$}, describing the joint appearance of agents and concepts, and henceforth the usage of the latter by the former, where each collaboration is a hyperlink $\hypere\in\mathcal{P}(\Aset{}\cup\Cset{})$.

\end{enumerate}

As such, an ``\emph{epistemic hypergraph}'' is properly defined by a triple $(\Aset{}, \Cset{}, \hyperE{})$. Dynamic epistemic hypergraphs are indexed with time, $\hyperE{t}$, and are considered to be growing: $t<t'\Rightarrow\hyperE{t}\subseteq\hyperE{t'}$. 

At each timestep, new teams are formed and thus hyperlinks appear, we denote this set by $\Delta\hyperE{t}$, such that $\hyperE{t}=\hyperE{t-1}\cup\Delta\hyperE{t}$. Note that $\Delta\hyperE{t}$ is not necessarily equal to $\hyperE{t}\setminus\hyperE{t-1}$ since some teams forming at $t$ may already have appeared in $\hyperE{t-1}$. 

See an illustration of this framework on Fig.~\ref{fig:epistemichypergraphtemporal}.

\medskip
We also define a projection operation 
for hyperlinks: given a hyperlink $\hypere\in\hyperE{t}$ and a subset $E\subseteq\Aset\cup\Cset$, the projection of $\hypere$ on the set $E$ is noted $\hypere^E=\hypere\cap E$.
For instance, the fact that all  hyperlinks contain at least one agent translates as $\forall \hypere$, $\Aproj{\hypere}\neq\emptyset$.

We can thus define a (dynamic) collaboration hypergraph $\{\Aproj{\hypere}\,|\,{\hypere\in\hyperE{t}}\}=\hyperA{t}\subseteq\mathcal{P}(\Aset{})$ whose hyperlinks connect team members, and a semantic hypergraph $\{\Cproj{\hypere}\,|\,{\hypere\in\hyperE{t}}\}=\hyperC{t}\subseteq\mathcal{P}(\Cset{})$ whose hyperlinks are sets of concepts mentioned in a given collaboration. 
In particular, $\hyperA{t}$ is isomorphic to a bipartite graph of collaboration, traditional in the literature \citep{newm:rand,guim:team}.

%


\subsubsection*{Neophytes and newcomers.}

We say that an agent $a$ is, at $t$, 
 a ``{\em neophyte in a given concept} $c\in\Cset{}$'' if s/he has never used $c$ at $t$: formally, if $\nexists\hypere\in\hyperE{t-1}, \{a,c\}\subseteq\hypere$. 
Otherwise, s/he is called an ``\emph{expert}''.

We say that an agent $a$ is a ``{\em newcomer}'' if 
s/he has never published before $t$, which is equivalent to say that $\nexists\hypere\in\hyperE{t-1}, a\in\hypere$.
Otherwise, s/he is called a ``\emph{veteran}''. 

Similarly, we say that a concept $c$ is a ``\emph{novelty}'' at $t$ if all agents are neophyte in this concept: 
$\nexists\hypere\in\hyperE{t-1}, c\in\hypere$.
Otherwise, it is a ``\emph{standard}''.

\subsubsection{Measures}\label{sec:measures}

\subsubsection*{Homogeneity of teams and expertise ratio.}

Given these basic concepts, we may first examine the composition of teams using a simple hypergraphic measure pertaining to the composition of teams in terms of a simple proportion of experts: ``how much are teams made of people familiar or not with a given concept which is used by the team?''. 

We call this proportion \emph{expertise ratio}, noted ``$\xi$''; for example, a paper on ``ants'' where half of the authors already worked on ants has a ratio of expertise in ``ants'' of $.5$.
Formally, the expertise ratio $\xi_{c,t}(\hypere)$ in concept $c\in\hypere^\Cset$ at time $t$ of team $\hypere$ is given by:
\[
\xi_{c,t}(\hypere)=\frac{|\{a\in\hypere^\Aset\;|\;\text{$a$ is an expert in $c$}\}|}
{|\{a\in\hypere^\Aset\}|}
\]

This notion, derived from the composition of a given team in terms of experts \hbox{vs.} neophytes in a given concept, expresses the socio-conceptual homogeneity of a team. See Fig.~\ref{fig:hyperexpert}.

\subsubsection*{Hypergraphic repetition.}

We may also express the degree of originality of the composition of a team and its subsequent groupings by measuring, in the broad sense, the proportion of already-existing associations of items, be it agents or concepts. More to the point, we may talk of \emph{social originality} by describing the rate of new associations of agents in a given team; or, dually, we will denote \emph{conceptual originality} by describing the proportion of new associations of concepts in a paper.\footnote{In which case, new concept associations are \emph{new} with respect to the whole system, consistently with the social case: \hbox{i.e.} this refers to concept associations which never existed in any paper of the preceding periods.}

\begin{figure}
\centering
{\sc Zebrafish}\\
\includegraphics[width=.99\linewidth]{\pics{}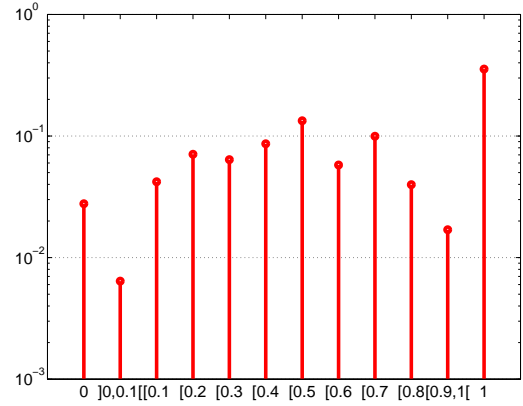}
\caption{\label{fig:teyobonly} Empirical distribution of the hypergraphic repetition rate for {\bf concepts, $r_t(\hypere^\Cset)$}.}
\end{figure}

More precisely, in the dyadic case, an interaction is said to be repeated if the two nodes already jointly appeared in a previous collaboration.
We extend this notion to the hypergraphic case: 
\begin{itemize}
\item We first say that a set of nodes has ``{\em previously co-occurred}'' if there is at least one previously-existing ($<t$) hyperlink including this set. We define the corresponding function $\rho_t$ as follows:
\[\rho_t(\hypere)=
\left\{\begin{array}{lll}
1&\text{if }\;\exists \hypere'\in\hyperE{t-1}, \hypere\subseteq\hypere'
\\
	0&\text{otherwise}.
	\end{array}\right.
\]

Thus, for instance, if $a$ and $a'$ never collaborated at $t$, we have $\rho_t(\{a,a'\})=0$.

\item The notion of hypergraphic repetition is properly defined for veteran agents and/or standard concepts --- by definition, repetition cannot occur with newcomers or novelties. 

Therefore, in the following formulas, hyperlinks $\hypere$ must be such that $\forall e\in\hypere$, $\exists\hypere'\in\hyperE{t-1}$ such that $e\in\hypere'$. 
In other words, we ensure the use of such hyperlinks by considering, $\forall \hypere\in\hyperE{t}$, truncated hyperlinks $\underline{\hypere}$ restrained to the set of previously-existing nodes, \hbox{i.e.}:
$$\underline{\hypere}=\hypere\cap \displaystyle\bigcup_{\hypere'\in\mathfrak{\hyperE{t-1}}}\hypere'$$

We then compute the \emph{hypergraphic rate of repetition for a hyperlink $\hypere\in\hyperE{t}$ as the proportion of subsets of this hyperlink that have previously co-occurred}:
\begin{align*}
r_t(\hypere)&=\frac{1}{2^{|\underline{\hypere}|}-|\underline{\hypere}|-1}\displaystyle\sum_{\substack{\hypere'\subseteq \underline{\hypere}\\|\hypere'|\geq 2}}\rho_t(\hypere')
\\
&=r_t(\underline{\hypere})
\end{align*}

Depending on the objectives, it might be appropriate to weight the relative importance of each subset of hyperlink $\hypere$ in the sums, for instance according to their size: for a discussion on weighting functions, see Appendix~\ref{app:weighting}.

\end{itemize}
Let us consider the following example: given a new collaboration $\hypere$ forming at $t$, $r_t(\hypere^\Cset)$ thus measures its hypergraphic concept repetition, \hbox{i.e.} how much the concepts of $\hypere^\Cset$ have been jointly associated, altogether, in previous periods. Eventually, we may plot the distribution of such values $r_t$ for all teams, as shown in Fig.~\ref{fig:teyobonly}. Put simply, it shows that about a third of teams have a hypergraphic conceptual repetition of $1$, \hbox{i.e.} all their concepts $\hypere^\Cset$ have already \emph{jointly} been used in the past.

\subsubsection{Estimating propensities of team formation}\label{sec:model}

\subsubsection*{Null-model of hypergraph.}

A null-model of new teams based on agents (resp. concepts) is defined such that, at each period $t$, 
we randomly create new teams respecting empirically-observed numbers of agents (resp. concepts) \emph{and} their respective numbers of team participations. What is fundamentally randomized is the exact composition  of teams in terms of who is collaborating with whom:  in our null-model, {\em team members are basically reshuffled}. Put differently, the null-model expresses the composition of teams as would be happening \emph{by chance}.

\medskip
In other words and more practically,
\begin{itemize}
\item we empirically measure:
	\begin{enumerate}
	\item the size of new teams appearing at $t$, \hbox{i.e.} the distribution of $|\hypere^\Aset|$ (resp. $|\hypere^\Cset|$) for $\hypere\in\Delta\hyperE{t}$,
	\item for every element $e\in\Aset$ (resp. $e\in\Cset$), the number of times it appears in newly-formed teams, \hbox{i.e.}: 
	\[\big|\{\hypere\in\Delta\hyperE{t}\,\text{ such that }\, \hypere\ni e\}\big|\]
	\end{enumerate}

\item we then generate an artificial, uniformly random set of new teams $\widetilde{\Delta\hyperE{t}}\subset\mathcal{P}(\Aset\cup\Cset)$ which respects above-mentioned distributions, that is:
	\begin{enumerate}
	\item same distribution of sizes of new hyperlinks,
	\item same distribution of participations of elements in these new hyperlinks.
	\end{enumerate}
\end{itemize}

In the remainder, we examine and compare the properties of the empirical $\Delta\hyperE{t}$ and the randomly-created $\widetilde{\Delta\hyperE{t}}$.

\subsubsection*{Propensity.}
In particular, we define the propensity of team formation with respect to a given function $f$ of a hyperlink (e.g. the hypergraphic rate of repetition) as, for each possible value $x$ of the function, the ratio between the observed number of new hyperlinks (events) $\hypere{}$ such that $f(\hypere{})=x$ and the randomly-created number of such events:
\begin{equation}
\Pi_t(x)=\frac{\big|\{\hypere\in\Delta\hyperE{t}\,\text{ such that }\, f(\hypere{})=x\}\big|}{\big|\{\hypere\in\widetilde{\Delta\hyperE{t}}\,\text{ such that }\, f(\hypere)=x\}\big|}
\end{equation}

Obviously, if this quantity is above 1 for a certain value of $x$, we say that this type of team empirically occurs more than expected; otherwise, less.

\section{Results}\label{sec:results}

We may now empirically appraise hypotheses 1-2-3 \& I-II.

\subsection{Simulation of the null-model}

We start by measuring the propensity of team formation, first with respect to simple expertise ratios and, second, with respect to  hypergraphic repetition rates. To this end, we simulate $2,500$ instances of above-defined null-model-based epistemic hypergraphs,  which are therefore {random} hypergraphs.\footnote{For reasons of computational complexity, we consider event sizes not greater than $10$ agents and $10$ concepts --- with this constraint we still consider no less than $89\%$ of the total original number of teams.} We then compare the composition of teams thus obtained with that of the empirical data.

\begin{figure*}[!t]
\begin{center}{\sc Zebrafish}\\
\begin{tabular}{|c|c|c|}\hline
\multirow{2}{1.75cm}{\em expertise ratio}&\sc observed&\sc theoretical\\\cline{2-3}
&\includegraphics[width=0.4\linewidth]{\pics{}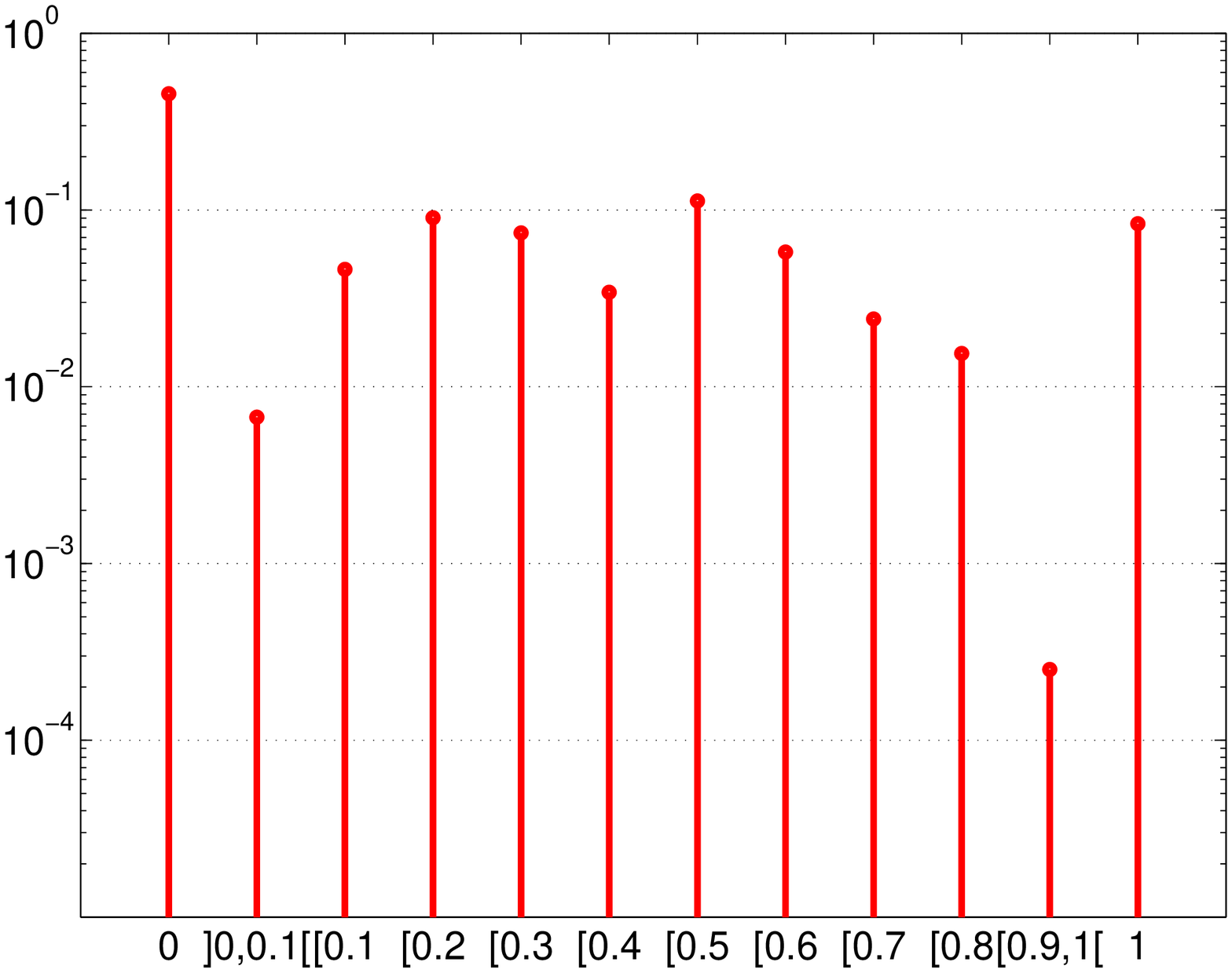}&
\includegraphics[width=0.4\linewidth]{\pics{}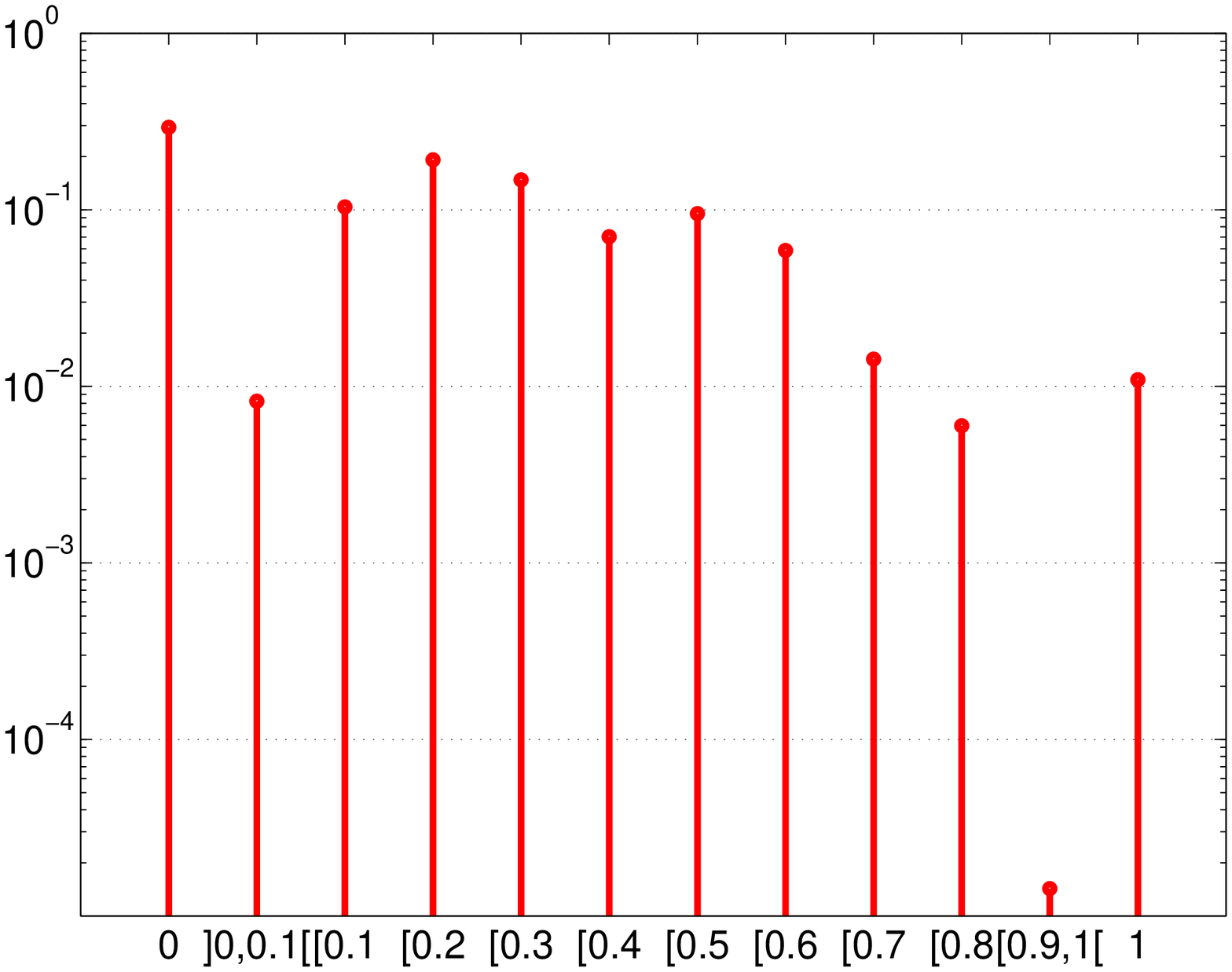}\\\hline
\end{tabular}
\caption{\label{fig:teyob}\label{fig:propex} Probability distribution of the expertise ratio on all teams aggregated over all years and all concepts (\textit{left:} observed, \textit{right:} theoretical). The computation of propensities below will be based on the ratio of such observed distributions over theoretical ones.
}
\end{center}
\end{figure*}

\begin{figure*}[!th]
\begin{center}
\begin{tabular}{cc}
	\begin{tabular}{c}{\sc Zebrafish}\\
	\includegraphics[width=.48\linewidth]{\pics{}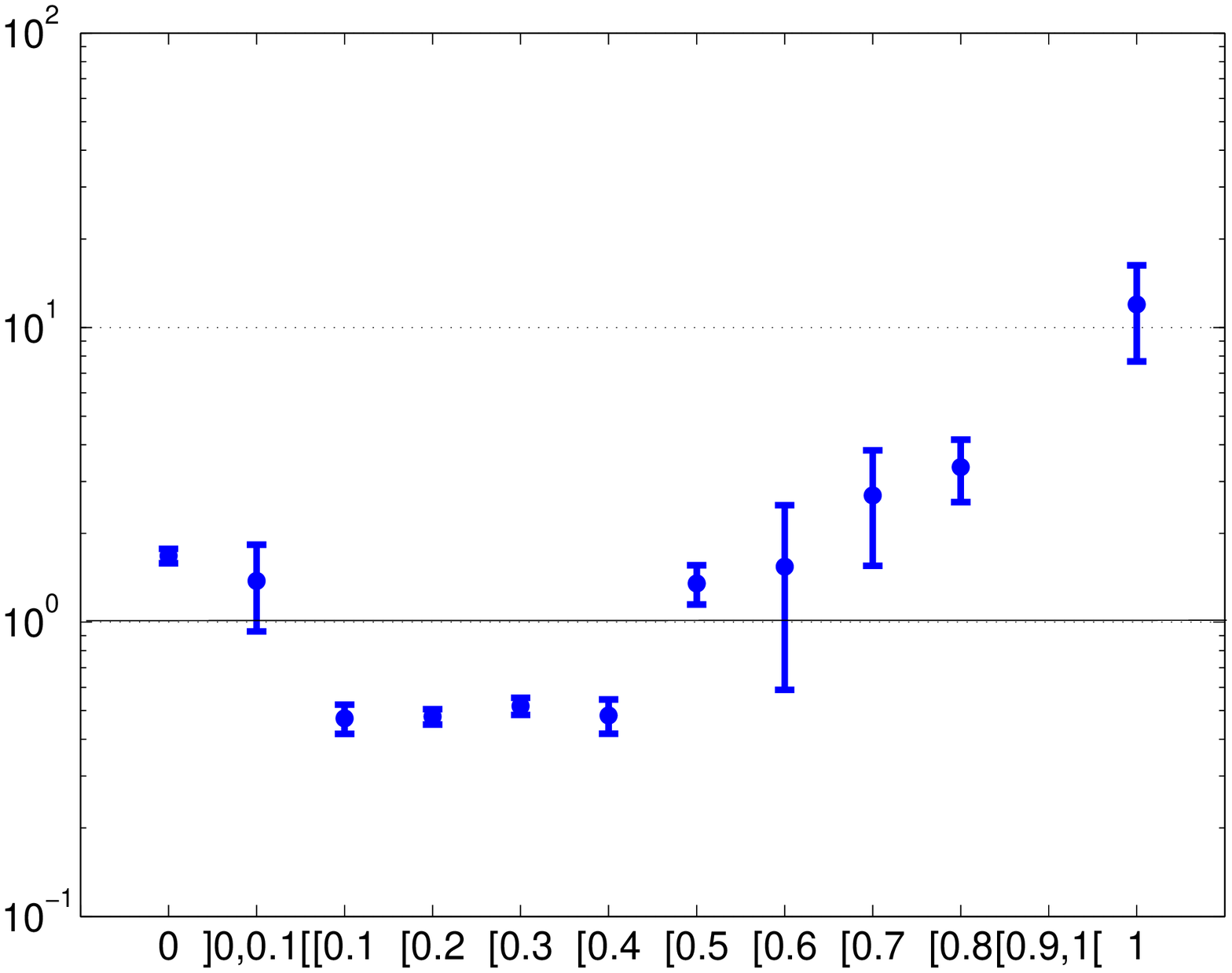}
	\end{tabular}
	&
	\begin{tabular}{c}{\sc Rabies}\\
	\includegraphics[width=.48\linewidth]{\pics{}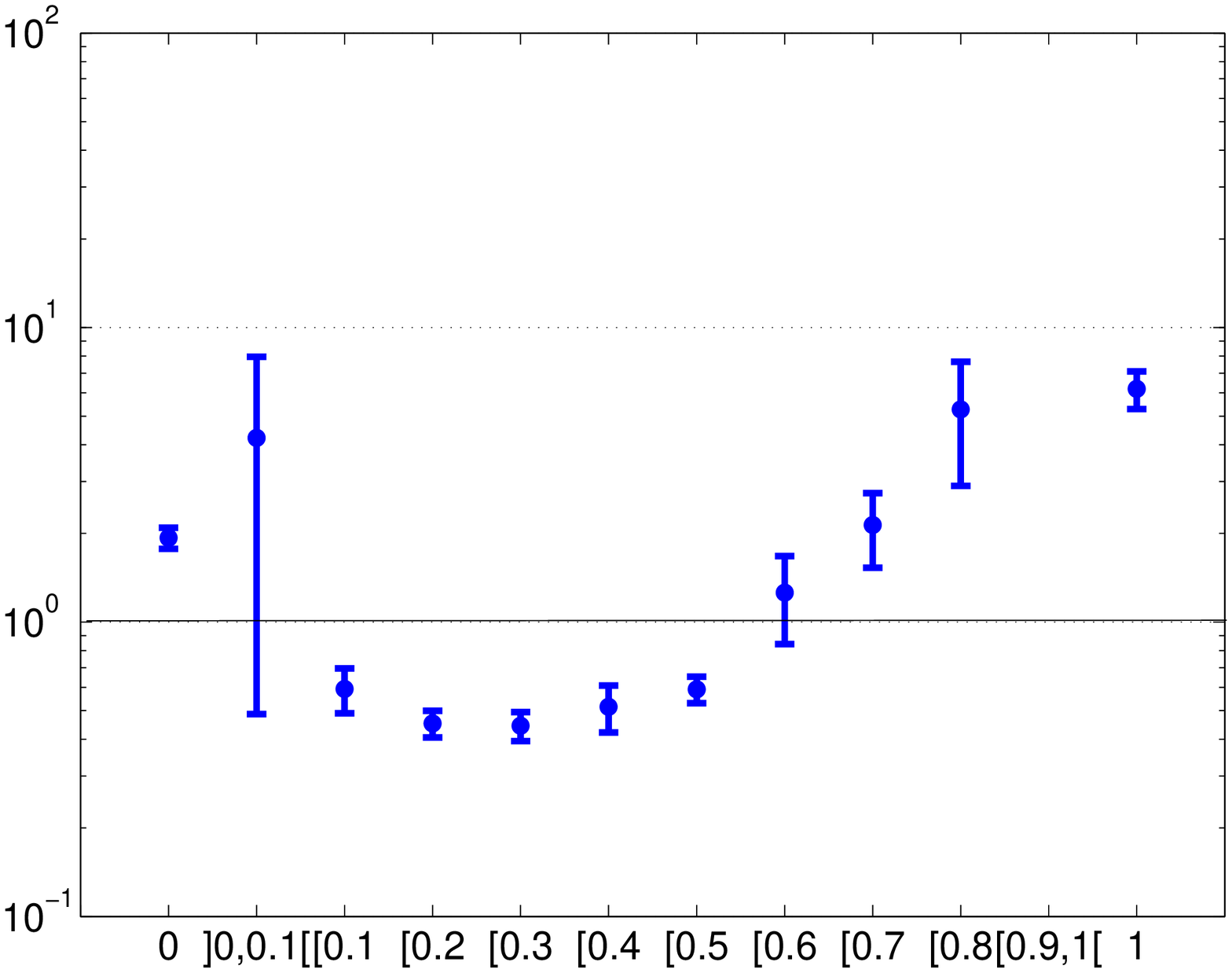}
	\end{tabular}
	\\
	\begin{tabular}{c}{\sc JECFA}\\
	\includegraphics[width=.48\linewidth]{\pics{}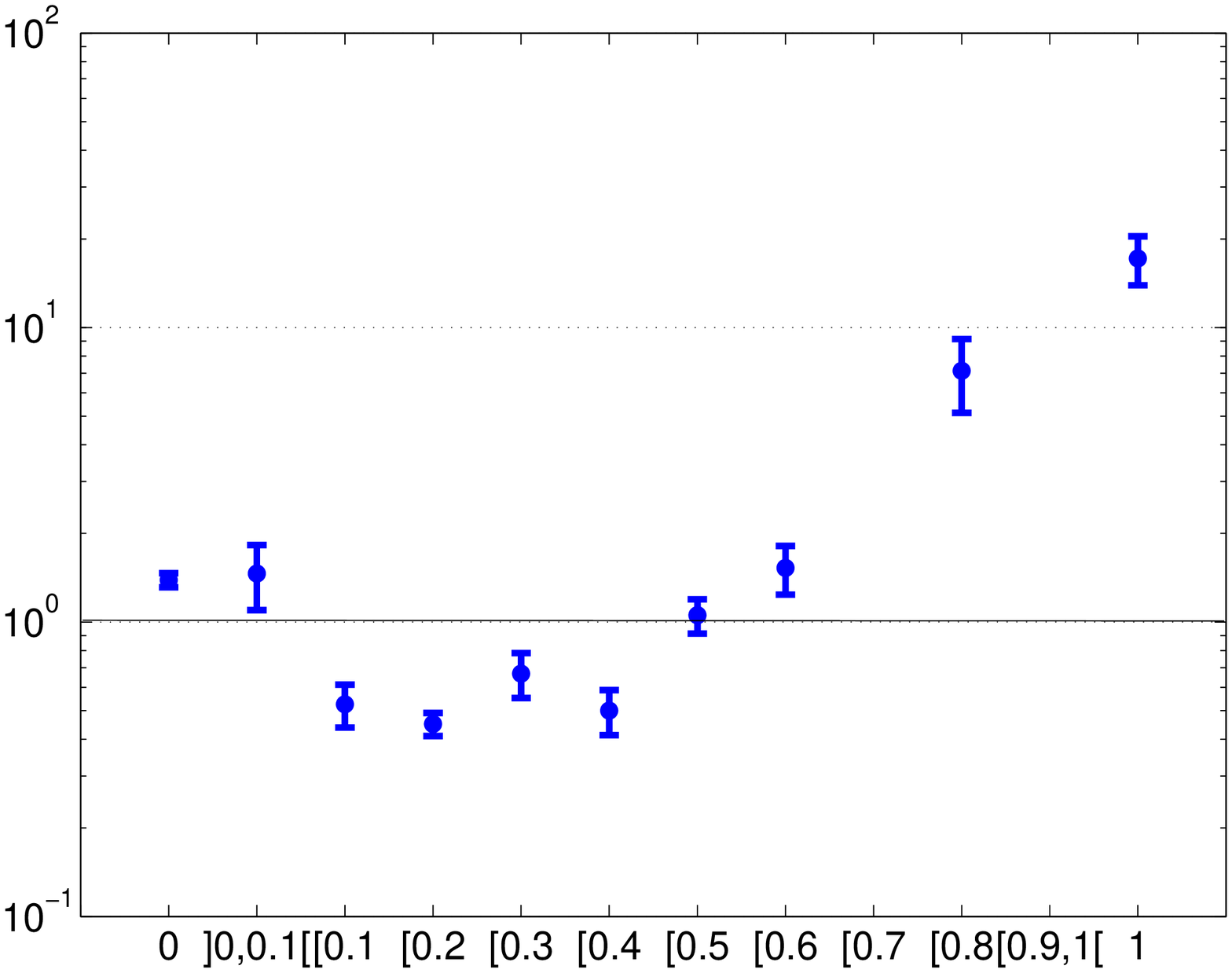}
	\end{tabular}
	&
	\begin{tabular}{c}{\sc JEMRA}\\
	\includegraphics[width=.48\linewidth]{\pics{}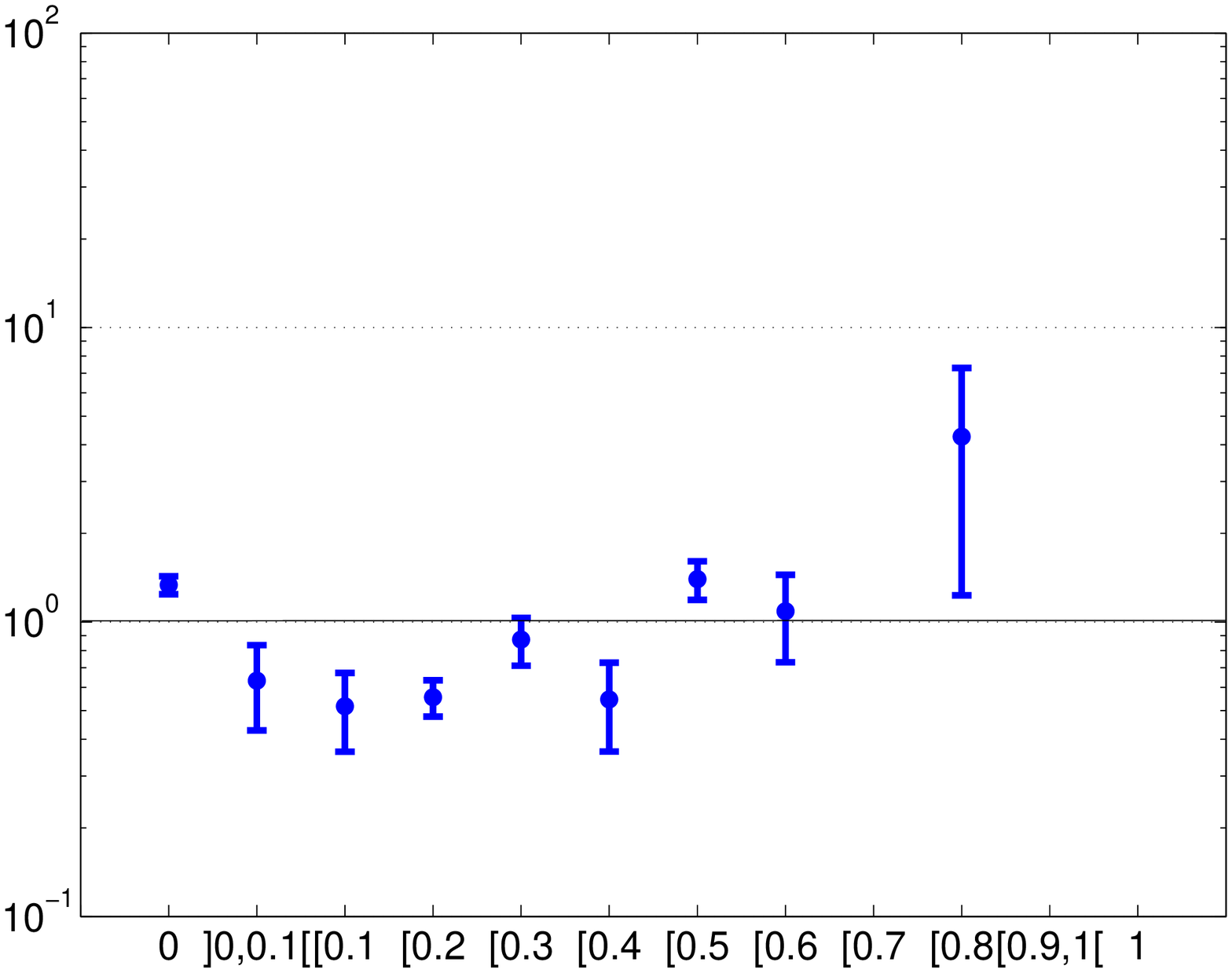}
	\end{tabular}

\end{tabular}
\caption{\label{fig:propreprop}Propensity for \textbf{proportions of experts per article}, from our real data vs expected from our random theoretical model --- averaged over all years, then over all concepts. {(Error bars correspond to $95$\% confidence intervals with respect to concept averages.)}}
\end{center}
\end{figure*}

\subsubsection*{Expertise ratio: socio-semantic homogeneity/heterogeneity}

Distinguishing agents who have already been associated with a concept (``experts'') and agents who are not yet associated (``neophytes''), we thus assess whether real teams involve agents of mixed backgrounds or not, relatively to a randomly-built set of teams. Details of this comparison are displayed on Fig.~\ref{fig:propex} for the \emph{zebrafish} case, which illustrates the composition of teams for various levels of expertise ratios, in both the real and random cases
. Corresponding propensities
, for both cases, are shown on Fig.~\ref{fig:propreprop}: their shapes are consistent across all datasets and consist of a U-shaped curve above $1$ for extreme values of expertise ratios (towards $0$ and towards $1$) and below $1$ for central values (typically, from $0.1$ to \hbox{ca.} $0.4--0.5$).

Empirically, we thus observe that there is a significantly high propensity of formation of teams composed of either experts only or newcomers only, with a significantly lower propensity for mixed teams. Teams involving a mixed proportion of experts and newcomers are thus less frequent than they should be. 




\subsubsection*{Hypergraphic rate of repetition: social or semantic homogeneity/heterogeneity}

Measuring now propensities of group formation with respect to hypergraphic rates of repetition, we can empirically exhibit the existence and influence of an implicit group structure which drives recurrent team formation --- this group structure exists along the two above-mentioned dimensions:

\begin{itemize}

\item {\em Social homogeneity/heterogeneity:}
With respect to agents, the hypergraphic rate of repetition measures the extent to which a team features repeated interactions among former collaborators. Once again, our results have to be compared to the null hypothesis for which teams are formed randomly. \hbox{Figure~\ref{fig:proprep}--\emph{top}} features the corresponding propensities which are several orders of magnitude higher than 1 for teams with a non-negligible proportion of such repetitions ($r>.1$)


\newcommand{\widthfig}{.37}
\begin{figure*}[!th]
\begin{center}
\begin{tabular}{cc}
	\begin{tabular}{c}
	{\sc Zebrafish}\\
	\includegraphics[width=\widthfig\linewidth]{\pics{}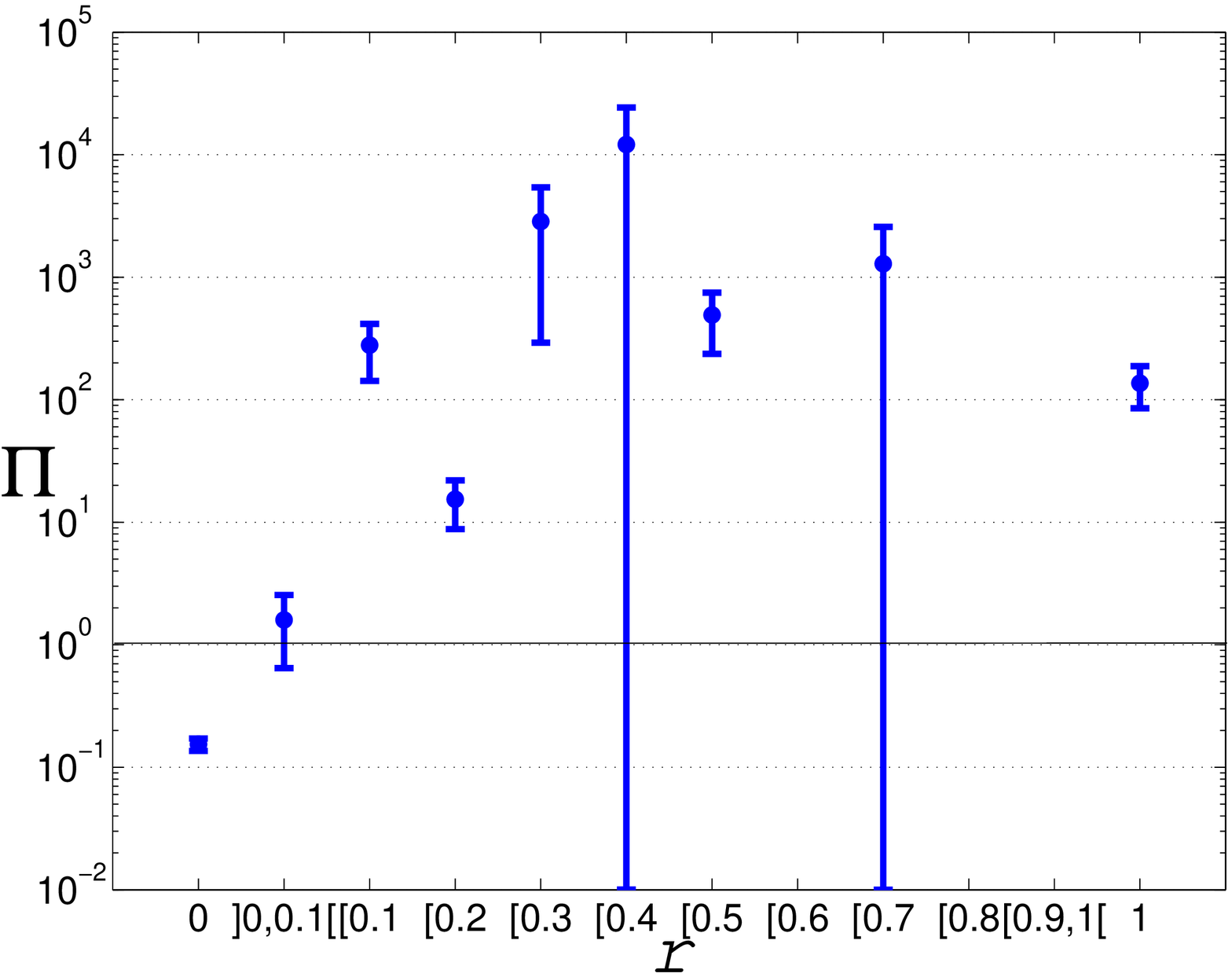}\end{tabular}
	&
	\begin{tabular}{c}
	{\sc Rabies}\\
	\includegraphics[width=\widthfig\linewidth]{\pics{}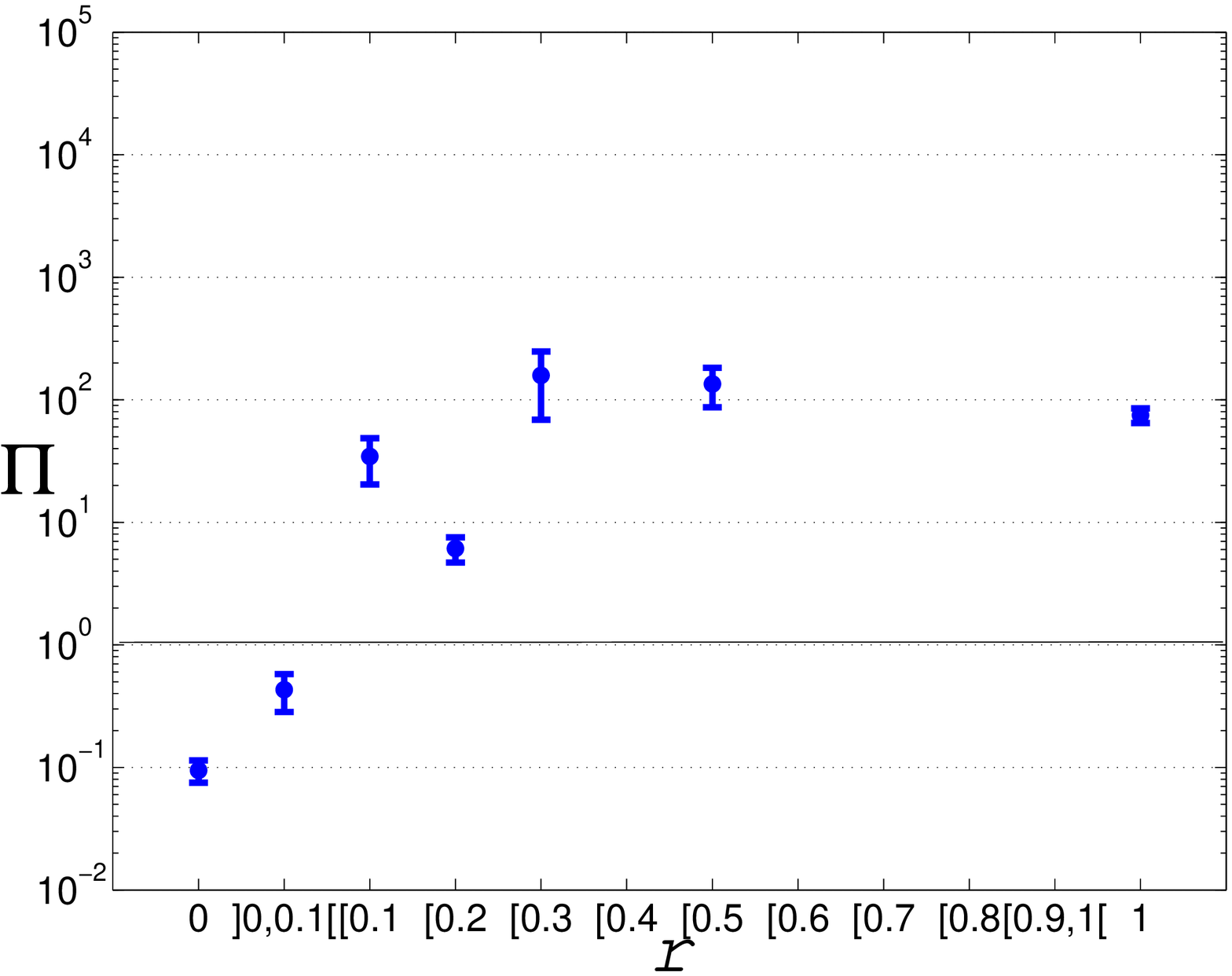}\end{tabular}
	\\
	\begin{tabular}{c}
	{\sc JECFA}\\
	\includegraphics[width=\widthfig\linewidth]{\pics{}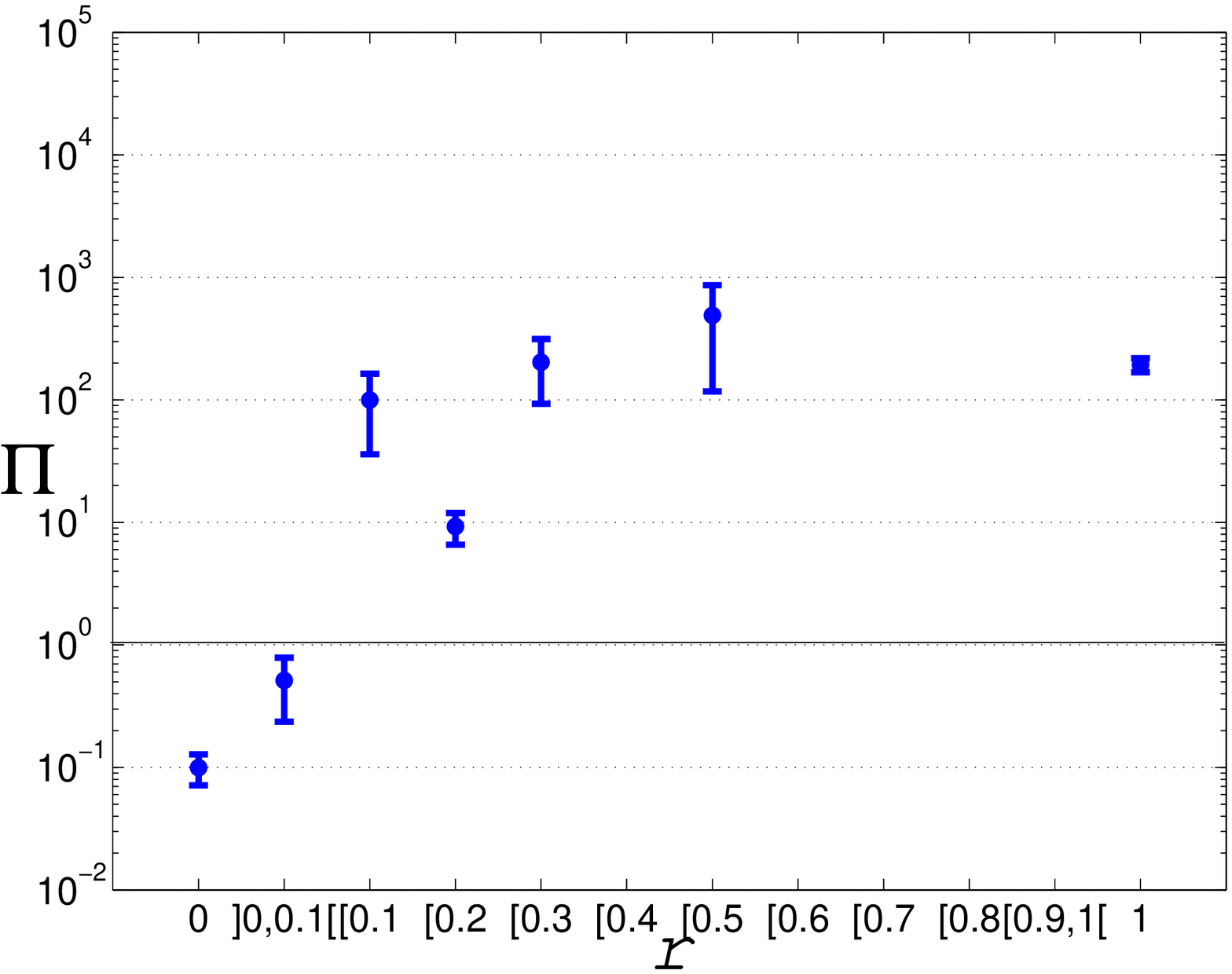}\end{tabular}
	&
	\begin{tabular}{c}
	{\sc JEMRA}\\
	\includegraphics[width=\widthfig\linewidth]{\pics{}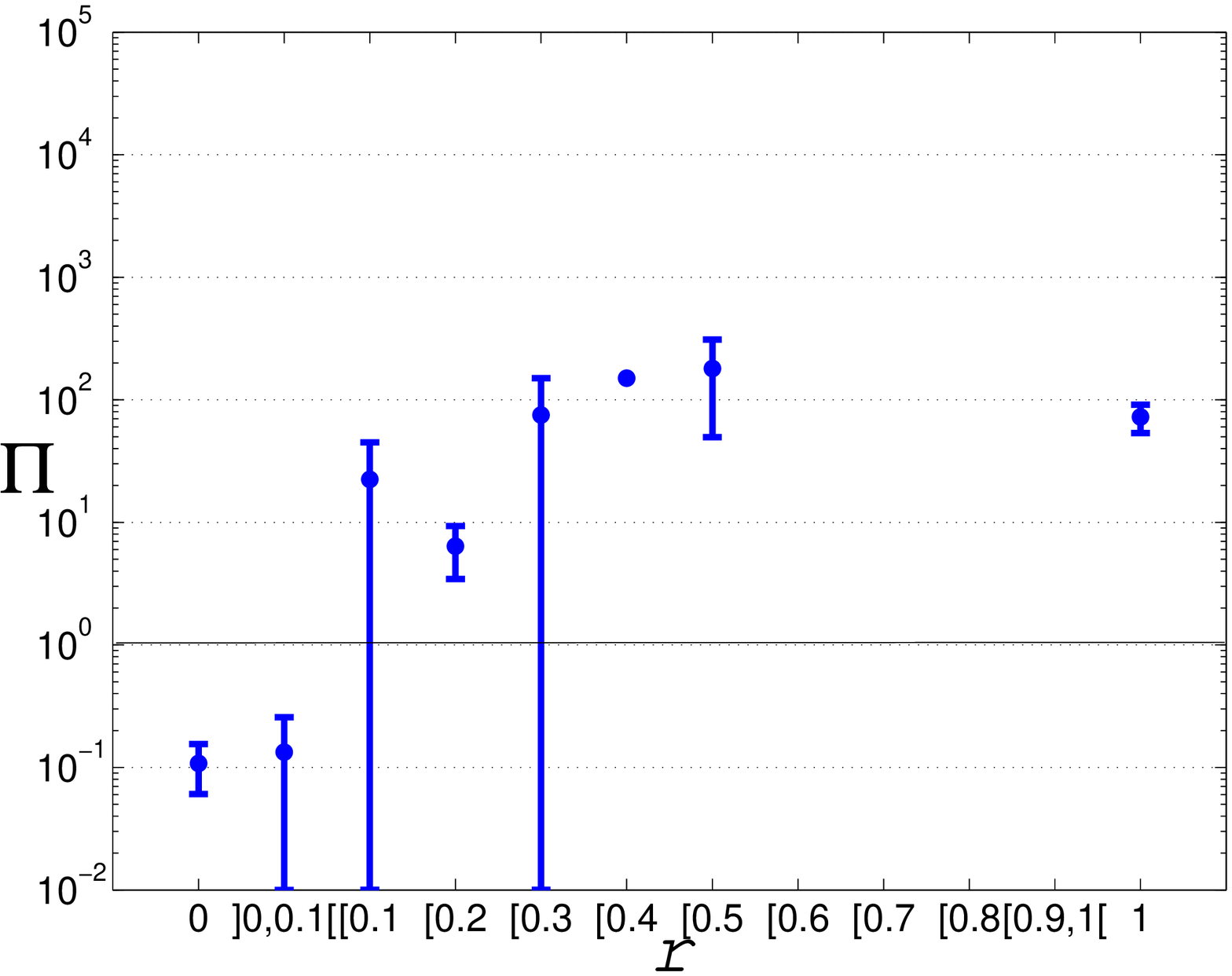}
	\end{tabular}
\end{tabular}
\end{center}
\hrule
\begin{center}
\begin{tabular}{cc}
	\begin{tabular}{c}
	{\sc Zebrafish}\\
	\includegraphics[width=\widthfig\linewidth]{\pics{}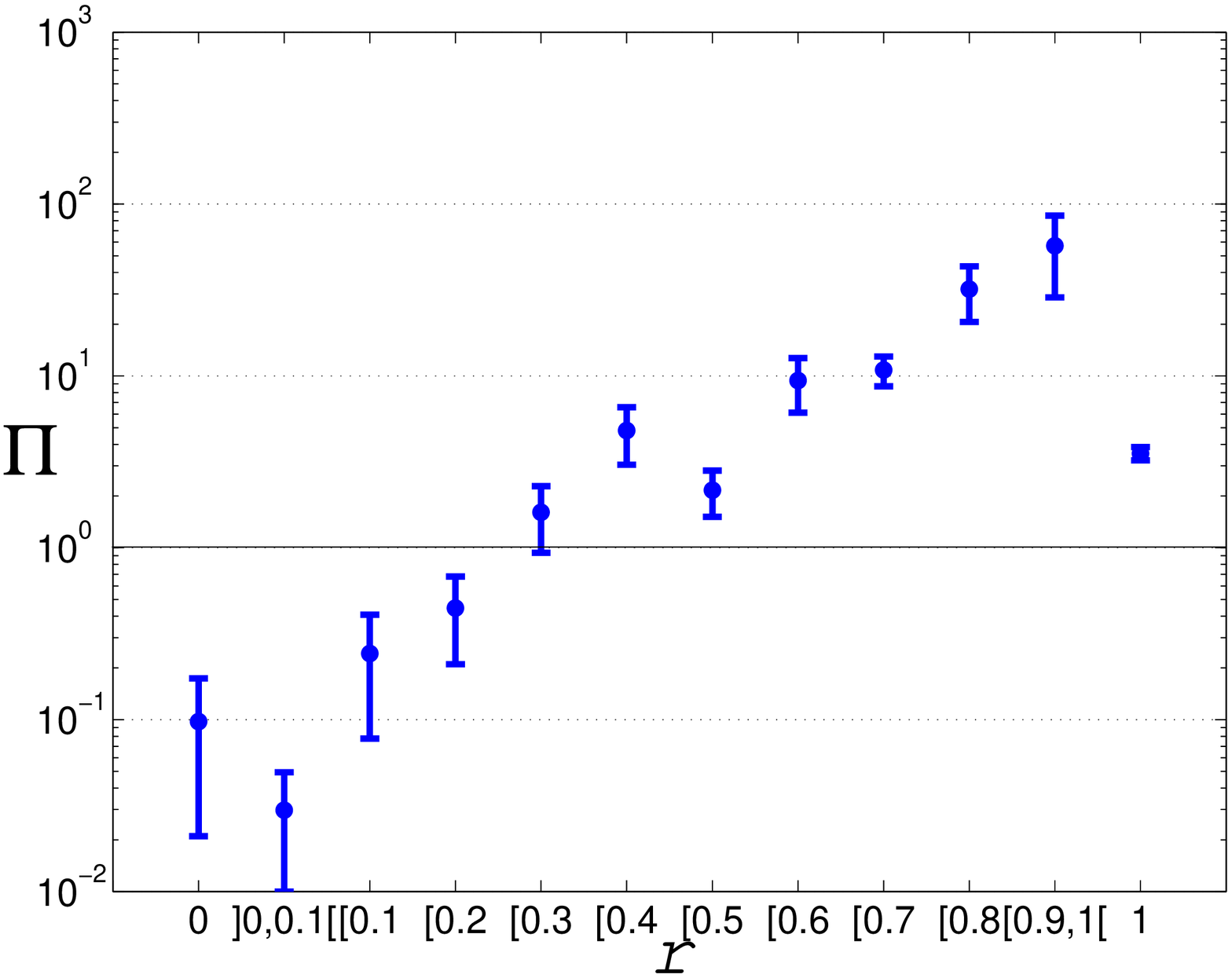}
	\end{tabular}
	&
	\begin{tabular}{c}
	{\sc Rabies}\\
	\includegraphics[width=\widthfig\linewidth]{\pics{}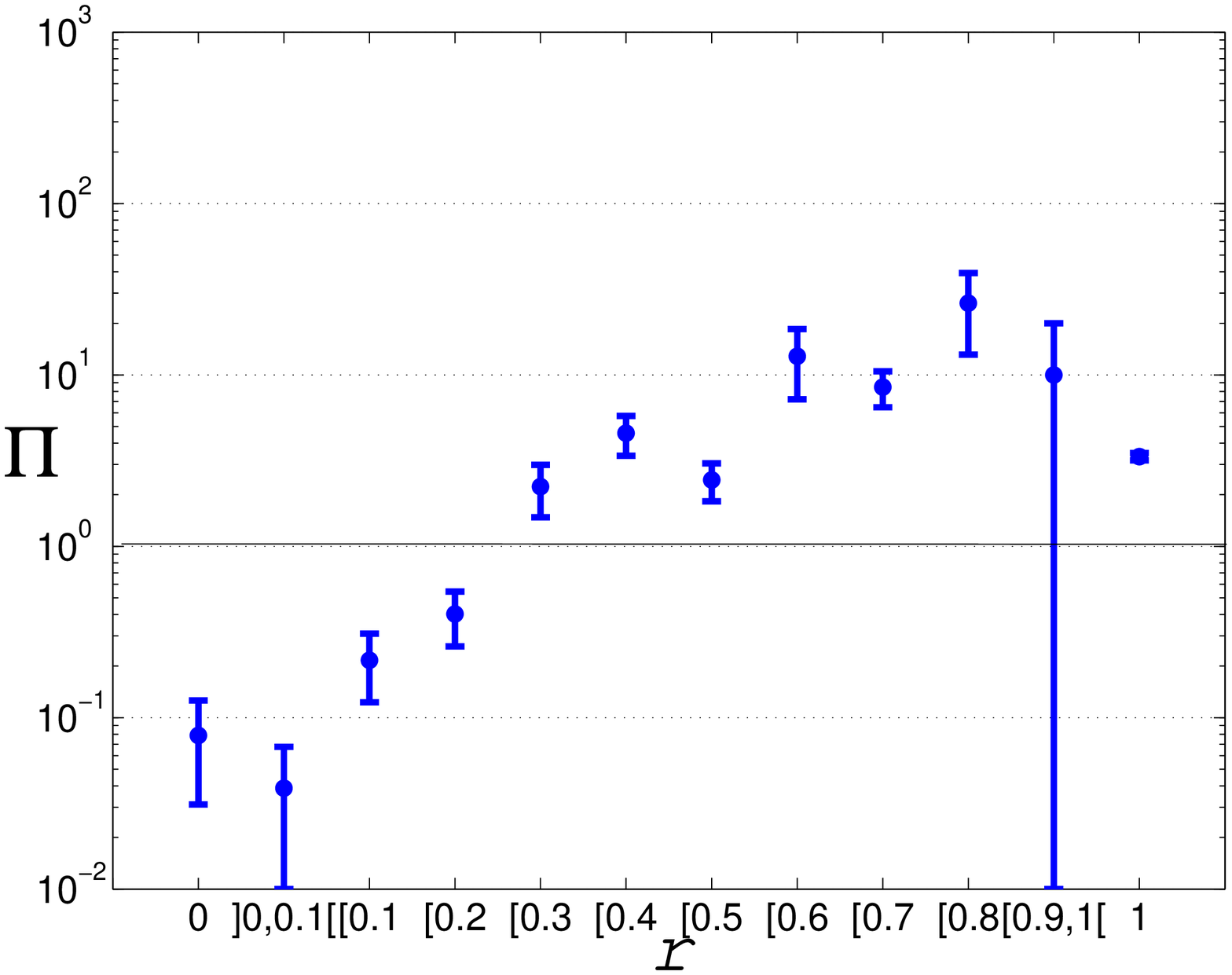}
	\end{tabular}
	\\
	\begin{tabular}{c}
	{\sc JECFA}\\
	\includegraphics[width=\widthfig\linewidth]{\pics{}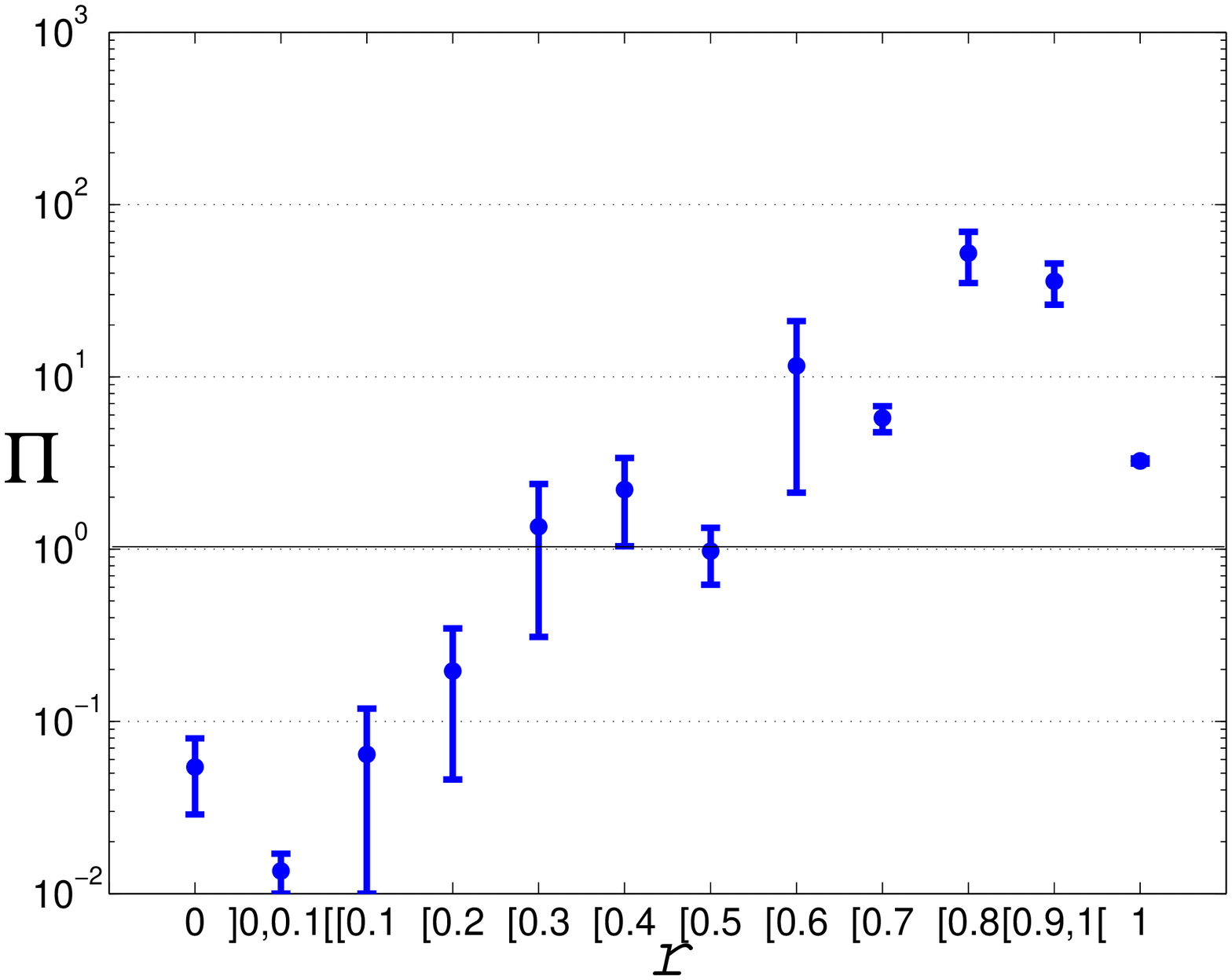}
	\end{tabular}
	&
	\begin{tabular}{c}
	{\sc JEMRA}\\
	\includegraphics[width=\widthfig\linewidth]{\pics{}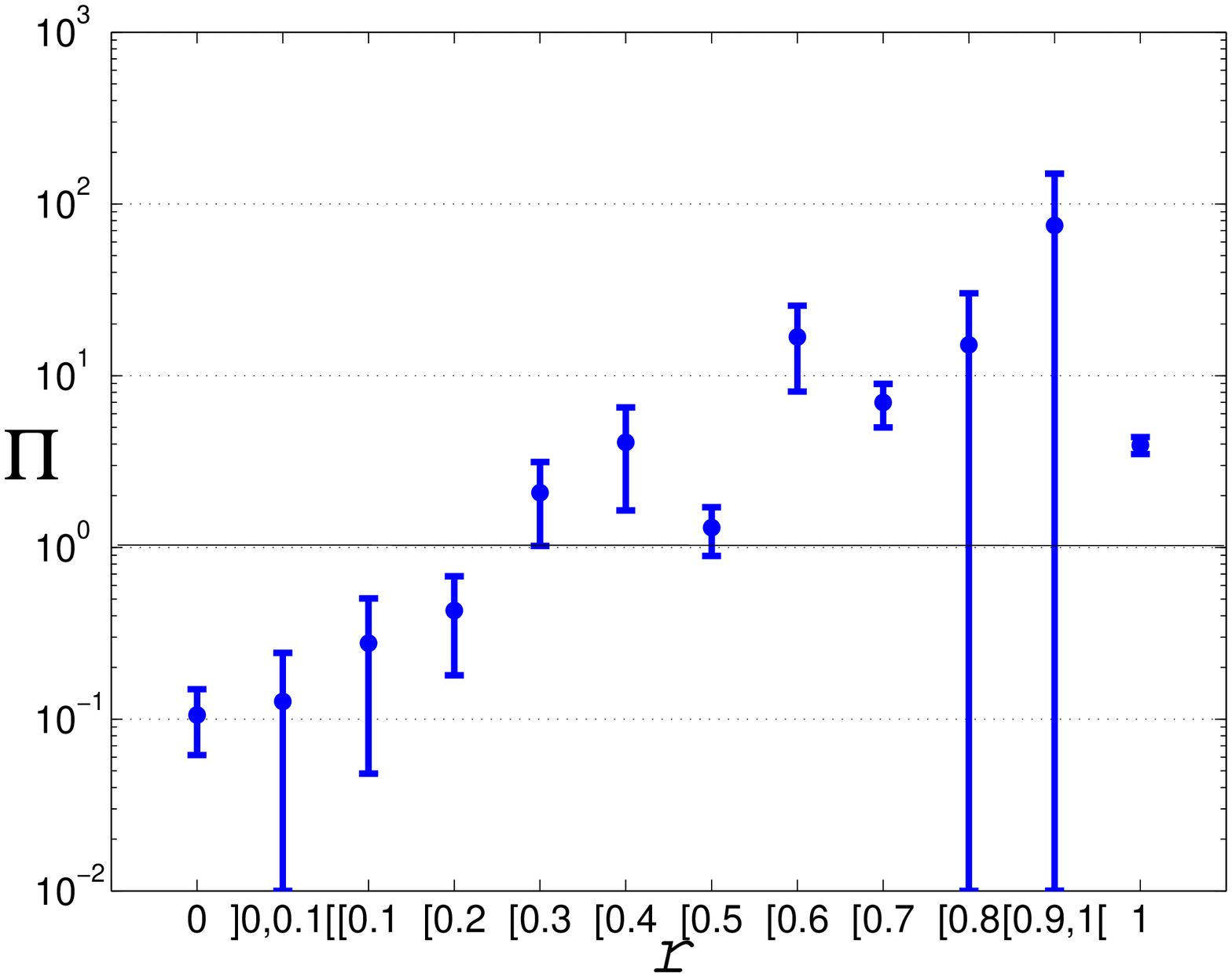}
	\end{tabular}
\end{tabular}
\caption{\label{fig:proprecon}\label{fig:proprep}
Propensity of team formation (random hypergraph \hbox{vs.} real data) with respect to hypergraphic repetition ratios for {\bf agents} (\emph{top}) and {\bf concepts} (\emph{bottom}). {(Values are averaged over all years, error bars correspond to $95$\% confidence intervals with respect to these averages.)}}
\end{center}
\end{figure*}

\item {\em Conceptual homogeneity/heterogeneity:}
Similarly, we measure the propensity of team formation with respect to repeated concept associations, addressing the following issue: ``are there cores of concepts which are likely to be recurrently associated, given that they were previously jointly used in previous papers?''
Results, shown on \hbox{Fig.~\ref{fig:proprecon}--\emph{bottom}}, demonstrate again (and even in a stronger fashion than in the social case) that there is a significant bias towards gathering \emph{groups of concepts} which were previously associated.
\end{itemize}

\subsection{Discussion of hypotheses}

It is now possible to review and check the afore-mentioned hypotheses. As follows from Fig.~\ref{fig:proprep}, it is clear that {\bf (H1)} and {\bf (H2)} are \emph{quantitatively} confirmed: teams with a high proportion of interaction repetitions or with a high proportion of repeated conceptual associations are much more likely than should be expected by chance.

\begin{figure}[!th]
\begin{center}
\includegraphics[width=1\linewidth]{\pics{}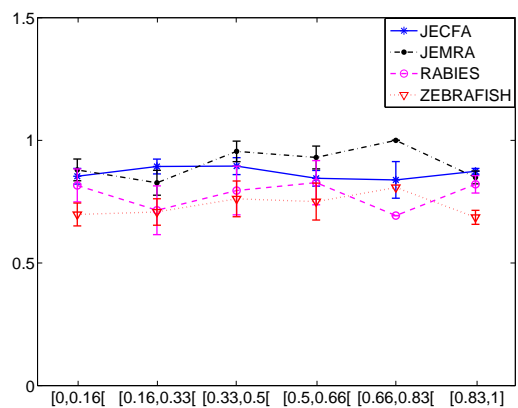}
\end{center}
\caption{\label{fig:correlation}Average semantic hypergraphic repetition ratio (\emph{y-axis}) for a given range of social hypergraphic repetition ratio (\emph{x-axis}).
{(Error bars correspond to $95$\% confidence intervals with respect to averages on each repetition ratio bin (in abscissa), such as \hbox{e.g.} $[0,0.1[$.)}}
\end{figure}

Additionally, and irrespective of the simulation model, we check if there is a correlation between semantic and social hypergraphic rates of repetition. As shown on Fig.~\ref{fig:correlation}, there seems to be no correlation between social and semantic originality in a collaboration (in our datasets, which come from varied backgrounds but are also focused on particular epistemic communities). This invalidates {\bf (H3)}: in other words, contrarily to intuition, new semantic associations do not stem more from original teams than from repeated teams. In other words, semantic innovation is as likely from agents who, globally, previously collaborated, as from new collaborations.\footnote{This does not mean, however, that the backgrounds of previous collaborators who are causing semantic innovation should necessarily be similar (semantic innovation might indeed come from repeated collaboration with individuals who have varied semantic backgrounds).}

As regards expertise, {\bf (HI)} --- ``teams gathering around a given topic should involve more individuals knowledgeable about it'' --- is partially confimed and partially contradicted by the empirical evidence. Firstly, teams with a high proportion of experts in a concept involved in the collaboration are much more likely, as shown on the right side of each graph on Fig.~\ref{fig:propreprop}, whose values are significantly above 1. \\
Yet, secondly, teams with a very small proportion of experts regarding a concept, \hbox{i.e.} high proportion of neophytes, are also significantly more likely, suggesting that part of the use of new concepts is also due to teams almost completely new to such concepts (even if, as is proved by (H1), these very teams are still more likely to stem from repeated collaborations). Put bluntly, new concept usage, and thus part of innovation, appears to stem both from teams significantly ignorant of such concepts and from teams globally knowledgeable about such concepts.

\begin{figure*}[!th]
\begin{center}
\begin{tabular}{cc}
	\begin{tabular}{c}{\sc Zebrafish}\\
	\includegraphics[width=.45\linewidth]{\pics{}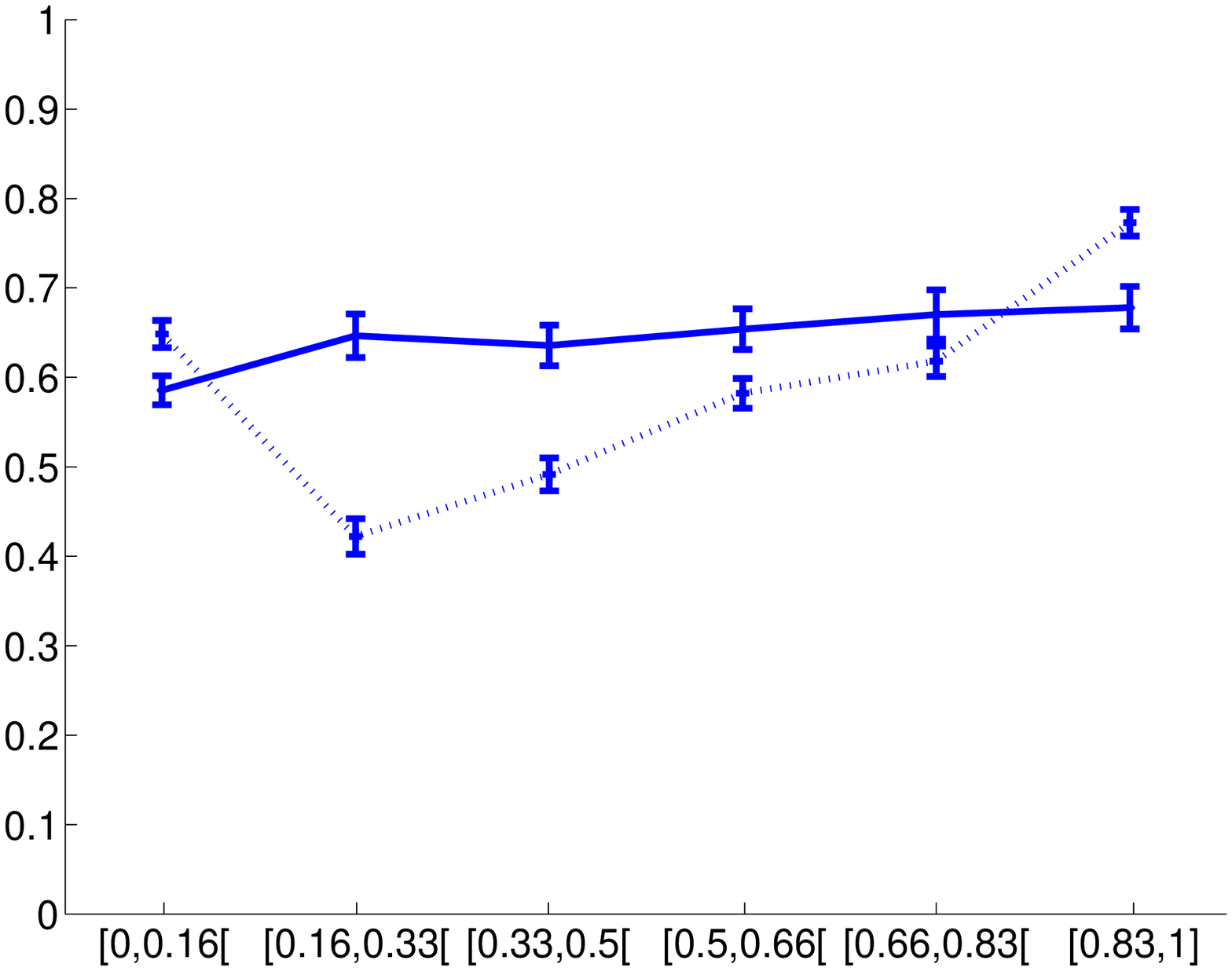}
	\end{tabular}
	&
	\begin{tabular}{c}{\sc Rabies}\\
	\includegraphics[width=.45\linewidth]{\pics{}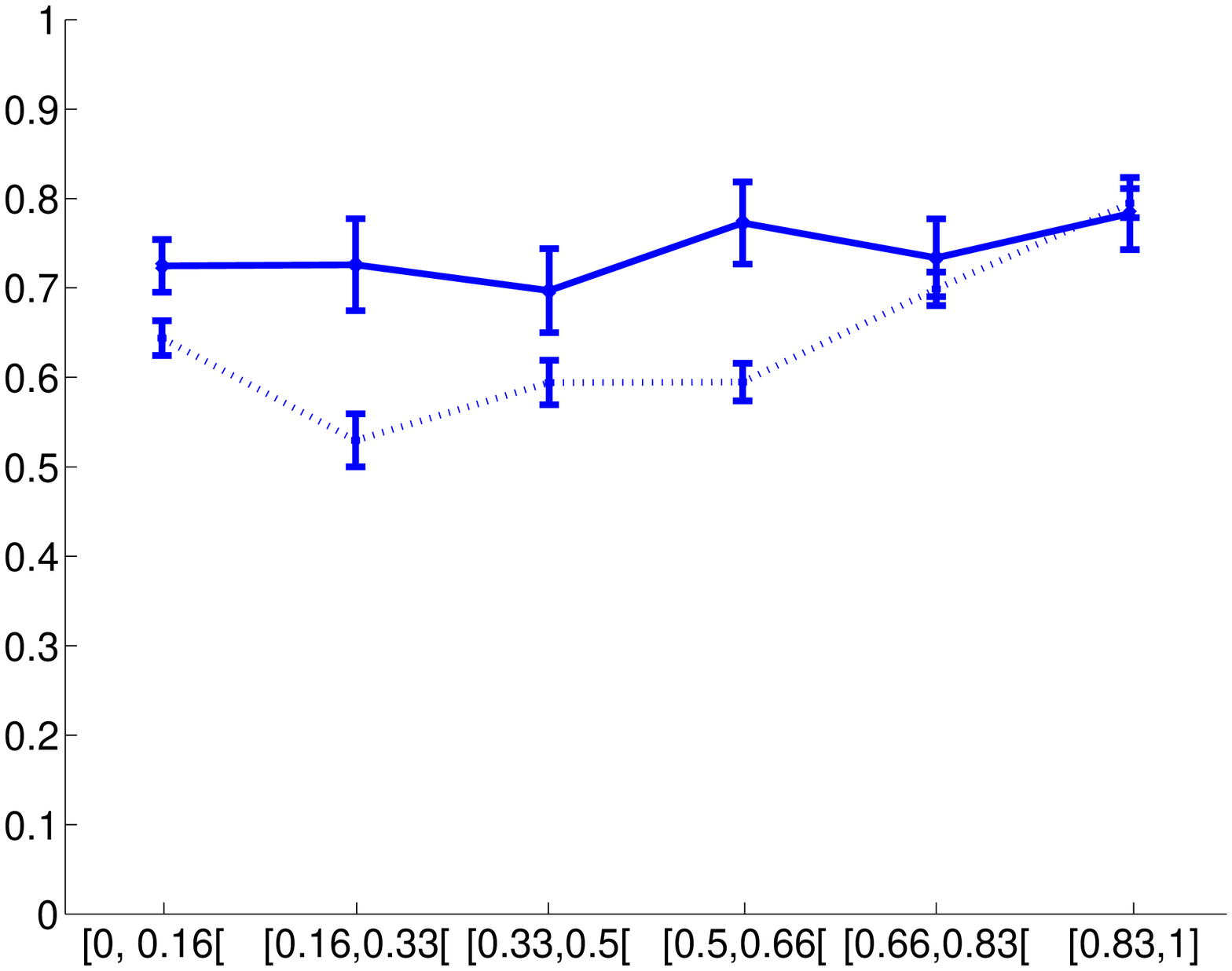}
	\end{tabular}
	\\
	\begin{tabular}{c}{\sc JECFA}\\
	\includegraphics[width=.45\linewidth]{\pics{}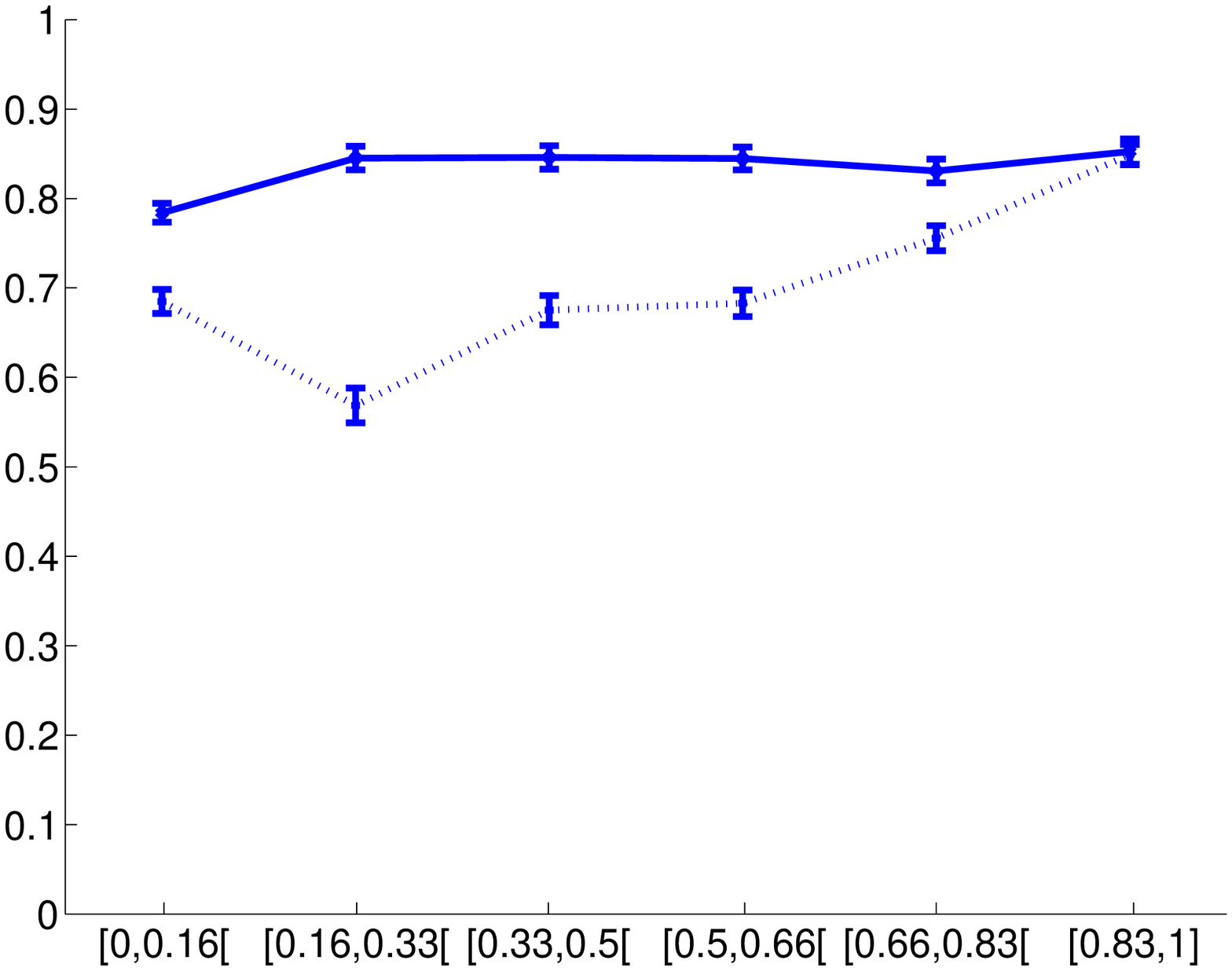}
	\end{tabular}
	&
	\begin{tabular}{c}{\sc JEMRA}\\
	\includegraphics[width=.45\linewidth]{\pics{}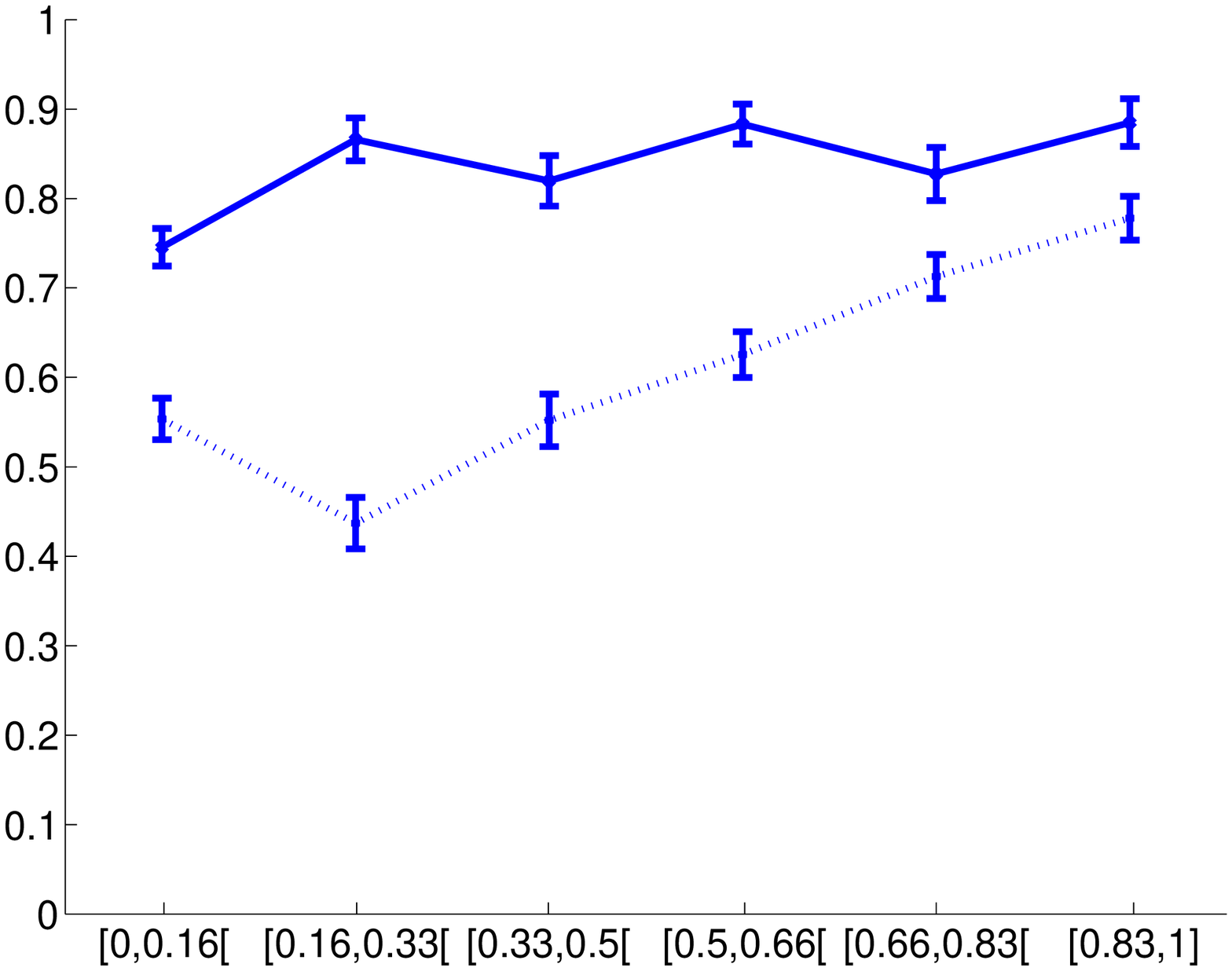}
	\end{tabular}
\end{tabular}
\caption{\label{fig:proprepexp} Average hypergraphic repetition ratios (\emph{y-axis}) with respect to expertise ratios (\emph{x-axis}): social (\emph{dashed line})  and semantic (\emph{plain line}) cases. {(Error bars correspond to $95$\% confidence intervals with respect to averages on each expertise ratio bin (in abscissa), such as \hbox{e.g.} $[0,0.1[$.)}}
\end{center}
\end{figure*}

From this observation that ``all-experts'' and ``all-neophytes'' teams are more likely, we may expect that such teams stem from underlying groups (either still working on the same topic, or working on a new topic, respectively) and thus have a higher social hypergraphic repetition ratio.  Similarly, those teams stemming from underlying groups are likely to carry normal, specialized science and have higher semantic hypergraphic repetition ratio (or lower originality). Figure~\ref{fig:proprepexp} sheds light on these issues by comparing average hypergraphic repetition ratios with expertise ratios. In particular, we observe that teams with a balanced composition of experts have a higher social originality (lower social hypergraphic repetition ratio), yet semantic originality remains constant across various values of expertise ratios. This partially confirms {\bf (HII)} as regards social originality and partially invalidates it as regards semantic originality: indeed, social originality is increased when there is a mixed proportion of experts, but not semantic originality.

\section{Concluding remarks}
We presented a formal  framework to appraise the underpinnings of collaboration formation with a hypergraphic  approach which encompasses both the meso-level of teams and the joint dynamics of social and semantic features. This allowed the quantitative estimation of the relative strength of social and semantic patterns behind academic team formation, by empirically studying several communities of scientists and estimating how the composition of teams, both cognitively and socially, diverges from a null hypothesis where collaborators and/or topics would be randomly chosen. 

We could thereby confirm several hypotheses as well as invalidate some hypotheses which had been established in a relatively qualitative fashion in the literature, or in a possibly misleading \emph{dyadic} form. More precisely, our measurements suggest a mechanism of team formation based on (i) a high likeliness to repeat previous collaborations patterns, not only dyadic but also $n$-adic interactions ($n\geq 3$) and (ii) a sensible confinement of groups of individuals, whose collaborations appear to depend largely on the history of team memberships, and, similarly, a sensible semantic confinement where associations of concepts depend largely on the repetition of previous associations. 
On the whole however, {the originality of a paper does not seem to stem from an original composition of the underlying team, while  a polarization appears between groups made of experts only or made of non-experts only, which altogether correspond to collectives exhibiting a high rate of repeated interactions.}

{\em Perspectives on models of academic collaboration.} Taking into account an implicit group structure, both at a social and at a socio-semantic level, as evidenced by the data, is likely to faithfully account for the structure of  academic collaboration networks. Indeed, the underlying low-level dynamics is plausibly closer to hypergraphic team formation mechanisms than would be allowed by a design based on dyadic interactions only.
As said before, this should not yield a lack of organizational thinking regarding the underpinnings of scientific production: beyond the step that constitutes our present contribution, an exhaustive approach about this type of collaboration mechanisms would indeed have to involve both epistemic hypergraphs and organizational features. In this respect, while we claim and show that hypergraphs make it possible to capture some interesting processes of team-based, knowledge-intensive production systems, we also emphasize that the richness of organizational mechanisms should not be shadowed by this formalism.

In line with our results, it should also be possible to determine which features, at the level-team, favor better collaborations --- not only in terms of semantic originality, but also in terms of quality and creativity of output, in a broad sense.

\paragraph{Acknowledgements.}
This work was partially supported by the Future and Emerging Technologies
programme FP7-COSI-ICT of the European Commission through project
QLectives (grant no.: 231200). We thank David Chavalarias and several anonymous reviewers for their useful comments.


\appendix
\section{Weighting functions}
\label{app:weighting}
A weighted hypergraphic repetition rate could be written as follows:
\[
r_t(\hypere)=\frac{\displaystyle\sum_{\substack{\hypere'\subseteq \hypere\\|\hypere'|\geq 2}}w_\hypere(|\hypere'|)\cdot \rho_t(\hypere')
}{\displaystyle\sum_{i\in\{2,...,|\hypere|\}}{w_\hypere(i){|\hypere|\choose i}}
}
\]
where $w_.$ is a weight function (given $\hypere$, $w_\hypere:\mathbb{N}\rightarrow\mathbb{R}$) which makes it possible to give more or less weight to particular subset sizes.

For instance:
\begin{itemize}
\item taking $w_\hypere(i)=1$, i.e. actually no weighting as has been used in the paper,
\[
r_t(\hypere)=\frac{1}{2^{|\hypere|}-|\hypere|-1}\displaystyle\sum_{\substack{\hypere'\subseteq \hypere\\|\hypere'|\geq 2}}\rho_t(\hypere')
\]
\item 
if instead $w_\hypere(i)=i$, i.e. weighting proportional to the size of the considered subset, \[r_t(\hypere)=\frac{1}{|\hypere|(2^{|\hypere|-1}-1)}\displaystyle\sum_{\substack{\hypere'\subseteq \hypere\\|\hypere'|\geq 2}}|\hypere'|\rho_t(\hypere')\]
\item if finally $w_\hypere(i)={|\hypere|\choose i}^{-1}$, i.e. weighting proportional to the number of possible subsets of size $|\hypere|$ in a set of size $i$, $$r_t(\hypere)=\sum_{\substack{\hypere'\subseteq \hypere\\|\hypere'|\geq 2}}\frac{\rho_t(\hypere')}{{|\hypere|\choose|\hypere'|}}$$
\end{itemize}
\end{document}